\documentclass[
 reprint,
 superscriptaddress,
 amsmath,amssymb,
 aps,
 floatfix
]{revtex4-2}

\usepackage{graphicx}
\usepackage{dcolumn}
\usepackage{bm}
\usepackage{physics}
\usepackage{csquotes}
\usepackage{mathtools}
\usepackage{amsfonts}
\usepackage{amssymb}
\usepackage{amsmath}
\usepackage[caption=false,labelformat=empty]{subfig}
\usepackage{diagbox}
\usepackage{dsfont}
\usepackage{hyperref}
\usepackage{placeins}
\usepackage{xspace}
\usepackage{nicematrix}
\NiceMatrixOptions{cell-space-limits = 2pt}
\hypersetup{linktocpage,colorlinks,citecolor={blue},pdfdisplaydoctitle=true,pdfpagemode=UseOutlines,bookmarksnumbered=true}
\usepackage{verbatim}
\usepackage{todonotes}
\usepackage{float}

\newcommand*{\gammain}{\ensuremath{\Gamma_{\text{in}}}}
\newcommand*{\id}{\ensuremath{\mathds{1}}}
\newcommand*{\tran}{\ensuremath{^T}}
\newcommand*{\inv}{\ensuremath{^{-1}}}
\newcommand*{\SU}{\ensuremath{\mathit{SU}}}

\newcommand{\mat}[1]{\mathbf{#1}}
\newcommand{\site}[1]{\ensuremath{\mathbf{#1}}}

\begin{document}

\title{ Gauged Gaussian PEPS - A High Dimensional Tensor Network Formulation for Lattice Gauge Theories}

\author{Ariel Kelman}
\affiliation{Racah Institute of Physics, The Hebrew University of Jerusalem, Givat Ram, Jerusalem 91904, Israel}
\author{Umberto Borla}
\affiliation{Racah Institute of Physics, The Hebrew University of Jerusalem, Givat Ram, Jerusalem 91904, Israel}
\author{Itay Gomelski}
\affiliation{Racah Institute of Physics, The Hebrew University of Jerusalem, Givat Ram, Jerusalem 91904, Israel}
\author{Jonathan Elyovich}
\affiliation{Racah Institute of Physics, The Hebrew University of Jerusalem, Givat Ram, Jerusalem 91904, Israel}
\author{Gertian Roose}
\affiliation{Racah Institute of Physics, The Hebrew University of Jerusalem, Givat Ram, Jerusalem 91904, Israel}
\author{Patrick Emonts}
\affiliation{Lorentz Institute, Leiden University, Niels Bohrweg 1, 2333 CA Leiden, Netherlands}
\author{Erez Zohar}
\affiliation{Racah Institute of Physics, The Hebrew University of Jerusalem, Givat Ram, Jerusalem 91904, Israel}

\date{\today}

\begin{abstract}
Gauge theories form the basis of our understanding of modern physics -- ranging from the description of quarks and gluons to effective models in condensed matter physics.
In the non-perturbative regime, gauge theories are conventionally treated discretely as lattice gauge theories.
The resulting systems are evaluated with path-integral based Monte Carlo methods.
These methods, however, can suffer from the sign problem and do not allow for a direct evaluation of real-time dynamics.
In this work, we present a unified and comprehensive framework for gauged gaussian Projected Entangled Pair States (PEPS), a variational ansatz based on tensor networks.
We review the construction of Hamiltonian lattice gauge theories, explain their similarities to PEPS, and detail the construction of the ansatz state.
The estimation of ground states is based on a variational Monte Carlo procedure with the PEPS as an ansatz state.
This sign-problem-free ansatz can be efficiently evaluated in any dimension  with arbitrary gauge groups, and can include dynamical fermionic matter, suggesting new options for the simulation of non-perturbative regimes of gauge theories, including QCD.
\end{abstract}
\maketitle

\tableofcontents

\section{Introduction \label{sec:introduction}}
Gauge theories are a central building block of modern physics. 
They are ubiquitous both in high-energy physics, forming the cornerstone of the Standard Model of particle physics \cite{peskin_introduction_1995}, and in condensed matter physics, where they appear for instance as effective low energy descriptions of quantum spin liquids~\cite{fradkin_field_2013, savary_quantum_2016-1}. 
In high energy physics, a gauge theory -- in the form of quantum chromodynamics (QCD) -- describes the interaction of quarks and gluons, the basic constituents of matter.
At high energies, QCD is asymptotically free~\cite{gross_asymptotically_1973} and can be described with  perturbation theory.
However, at low energies, the coupling constants of QCD are large, rendering a perturbative expansion of the theory impossible. 
This regime is especially interesting since it describes processes such as confinement and hadronisation, the formation of protons and neutrons out of quarks.
The non-perturbative character of the low-energy regime makes the study of confinement especially difficult.
It is within this context that our work is relevant.

One successful approach to non-perturbative gauge theories is lattice gauge theories (LGTs)~\cite{wilson_confinement_1974, kogut_hamiltonian_1975, kogut_introduction_1979}. 
In this approach, space (or spacetime) is discretized, i.e. placed on a lattice.
The resulting discretized gauge theory is usually formulated in Euclidean spacetime, i.e. time is considered as an additional discrete coordinate, and calculations are done with the path integral formalism.
This action-based formulation resembles a statistical physics problem~\cite{kogut_introduction_1979} and is amenable to Monte-Carlo computations~\cite{creutz_monte_1980, creutz_monte_1983}. 
Path-integral based Monte Carlo methods are the current gold standard for computing the hadronic spectrum for comparison with experimental data~\cite{aoki_flag_2021}. 
However, this approach suffers from two very significant issues: 
(a) working in Euclidean spacetime, i.e. imaginary time, prevents the direct study of real-time evolution; (b) the sign problem, which arises when quantities intended for use as probabilities in Monte-Carlo sampling are not guaranteed to form a valid probability distribution (the ``probability", often an exponential of the action, may be negative or complex)~\cite{troyer_computational_2005}.
If a system suffers from the sign-problem, Monte Carlo computations converge exponentially slowly;  large parts of the QCD phase diagram therefore remain unexplored \cite{fukushima_phase_2011}.

Keeping the discretized approach to space but working with continuous time, a Hamiltonian formulation is an alternative to the action-based formulation in which both space and time are discretized~\cite{kogut_hamiltonian_1975}.
Because time is continuous and an explicit variable in this formalism,  problem (a) is solved automatically.
However, a Hamiltonian formulation requires the explicit definition of a Hilbert space, which can be infinite dimensional (even for systems of finite size), depending on the gauge theory.
In this case, or if an exact solution is not possible due to the system size, variational methods are a good alternative.
The main challenge in a variational approach is the definition of a good ansatz state that captures the relevant characteristics of the problem at hand.

A versatile class of ansatz state are tensor networks, which are states constructed out of local tensors, which are ``contracted" to produce states of the system of interest. The local tensors have extra indices (these are the ones that are ``contracted", i.e. traced out), which correspond to extra virtual (i.e. non-physical) degrees of freedom that are used to define interactions between tensors associated with neighboring lattice sites.
They provide a powerful and conceptually compelling way of representing quantum states based on their entanglement properties. 
In addition to their theoretical importance, they have proven to be an invaluable numerical tool as tensor network based algorithms often scale polynomially, rather than exponentially, with system size~\cite{white_density_1992, fannes_finitely_1992, schollwock_density-matrix_2011, cirac_matrix_2021}. 
These are particularly suitable for finding ground states of gapped local Hamiltonians~\cite{hastings_area_2007,white_density_1992, fannes_finitely_1992}, as well as investigating time evolution~\cite{daley_time-dependent_2004, zwolak_mixed-state_2004}, and thermal states~\cite{verstraete_matrix_2004}.

Much of the success of tensor networks stems from the fact that one-dimensional tensor networks, known as matrix product states (MPS), can be contracted efficiently, i.e. norms and expectation values can be evaluated efficiently~\cite{cirac_matrix_2021}.
In higher dimensions this is no longer the case, and the contraction of two-dimensional tensor networks, known as projected entangled pair states (PEPS)~\cite{jordan_classical_2008,cirac_renormalization_2009,orus_practical_2014,cirac_matrix_2021}, scales exponentially with the system size~\cite{schuch_computational_2007}.
To alleviate this problem, infinite PEPS (iPEPS) are commonly used for translationally invariant systems~\cite{corboz_simulation_2010-1}.
 
Given their usefulness in condensed matter, tensor networks have been successfully applied to LGTs.
In one space dimension, MPS both reproduce known results and overcome bottlenecks of conventional methods, such as the sign problem and finite chemical potential problems -- see, e.g., the earlier work~\cite{banuls_mass_2013,buyens_matrix_2014,rico_tensor_2014,kuhn_non-abelian_2015,banuls_thermal_2015,pichler_real-time_2016,buyens_hamiltonian_2016} as well as the review papers~\cite{dalmonte_lattice_2016,banuls_simulating_2020,banuls_review_2020}. 
As a first step beyond one dimensional problems, one can consider small, compact, extra dimensions from the one-dimensional case and use MPS for the computation~\cite{tschirsich_phase_2019,gonzalez-cuadra_robust_2020,borla_quantum_2022,brenig_spinless_2022}.
In more general higher dimensional settings, LGTs have been studied numerically using tensor networks, for example using tree tensor networks and infinite PEPS methods~\cite{tagliacozzo_entanglement_2011,tagliacozzo_tensor_2014,crone_detecting_2020,robaina_simulating_2021,felser_two-dimensional_2020,magnifico_lattice_2021,montangero_loop-free_2022,cataldi_21d_2023}.
A complementary approach is that of the tensor renormalization group (see, e.g., the review~\cite{meurice_tensor_2022}), but this is not a Hamiltonian method, but rather one in which calculations are performed within a Euclidean spacetime and an action formalism.

In this work, we will review a tensor-network based approach for higher dimensions based on Gauged Gaussian (fermionic) PEPS (GGFPEPS).
The idea of a gauged Gaussian state is expected to work well for LGTs since the relevant Hamiltonians approach the free (Gaussian) Hamiltonian in the continuum limit.
Furthermore, the gauging of the state structurally resembles the minimal coupling procedure used to lift the free theory to an interacting one~\cite{emonts_gauss_2020}.
Practically speaking, the power of this approach resides in the fact that for a given fixed configuration of the gauge fields, the ansatz state is Gaussian and expectation values of relevant observables can be computed efficiently using the covariance matrix formalism~\cite{bravyi_lagrangian_2005}. 
The Gaussian nature of the state allows efficient contractions of the tensor networks in higher dimensions, because the state can be fully captured by its covariance matrix, which is efficiently computable.
The full state is taken as an integral over gauge field configurations, and is not Gaussian -- providing reason to think it may be expressive enough to include ground states of interacting theories.
As seen through some toy model constructions, showing some relevant physics qualitatively~\cite{zohar_fermionic_2015,zohar_projected_2016}, it turns out that these states capture the ground state of lattice gauge theories in high dimensions~\cite{zohar_combining_2018, emonts_variational_2020}.

The GGFPEPS are the ansatz state for a variational Monte Carlo procedure~\cite{sorella_generalized_2001, sorella_wave_2005}.
In the presence of dynamical gauge fields, expectation values of observables are extracted by averaging over all possible gauge configurations, in a way similar to path integration. 
While the latter is a formidable task in principle, Monte-Carlo techniques make this problem tractable~\cite{zohar_combining_2018, emonts_variational_2020}. 
The probability needed to perform Monte-Carlo sampling depends only on the norms of quantum states, and thus is guaranteed to provide a valid probability distribution, thereby avoiding the sign problem~\cite{troyer_computational_2005}.
This approach has been successfully applied to ground state searches for pure gauge $\mathbb{Z}_2$ and $\mathbb{Z}_3$ LGTs~\cite{emonts_variational_2020, emonts_finding_2023}.
The idea of using variational Monte Carlo for lattice gauge theories is not restricted to GGFPEPS; other constructions such as non-Gaussian states or neural network-based states show promising results as well~\cite{bender_real-time_2020,luo_gauge_2020,luo_gauge_2022, bender_variational_2023}.

This paper unifies previous work spread across various papers, presents the formalism in a general and self-contained form, and adds details regarding both analytics and numerics which were not discussed in previous work~\cite{zohar_combining_2018,emonts_variational_2020,emonts_finding_2023}. 
It is supposed to serve as a self-contained guide to the topic, and motivate the gauged Gaussian PEPS ansatz. 
We review the basics of building Gaussian fermionic PEPS, focusing on the physical scenarios relevant to gauging matter, and then introduce a generalized gauging procedure for these states which works for arbitrary gauge groups. We introduce a new consistent way of separating the parts of the ansatz which are coupled to the matter and decoupled from it (which we refer to as different types of fermions), add fermionic flavors and chemical potentials, and discuss the choice of the initial gauge field configuration before coupling to the matter.
We conclude by sketching the algorithm used in our computations, and mention some manipulations which greatly reduce the computational load of numerical calculations. In particular, we introduce a way to reduce the Monte Carlo integration space by making use of gauge fixing, and provide details on the separation of our ansatz into layers, as used in~\cite{emonts_finding_2023}.

Throughout this work, summation over repeated indices is assumed, unless they are bracketed, correspond to coordinates, label irreducible representations, or otherwise noted.

\section{The Physical System\label{sec:phys_system}}

\begin{figure}[t]
    \hspace{-10pt}
	\includegraphics[width=0.7\linewidth]{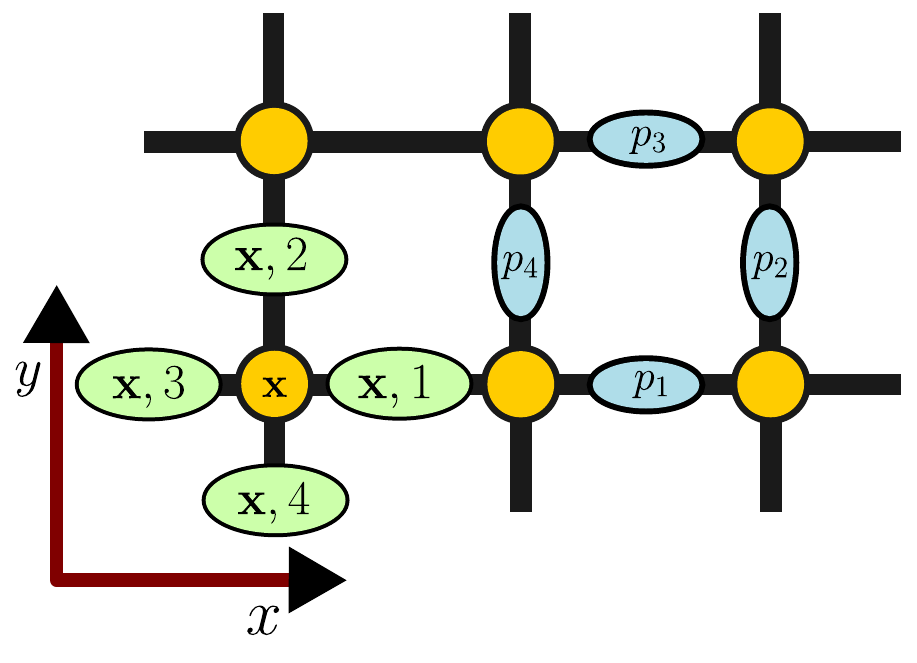}
	\caption{Representation of the lattice and of the labeling convention used for sites and links. When using a specific site $\site{x}$ as a reference, the links emanating out of it are labeled as in the green ovals on the left. When referring to a plaquette $p$, the links around it are labeled as in the blue ovals on the right.}
    \label{fig:lattice_scheme}
\end{figure}

We consider gauge theories on a spatial square lattice,
with dimension $d=2$ or $d=3$, with periodic boundary conditions. 
While the physics can differ significantly with other boundary conditions, the construction of the ansatz states discussed below may be generalized in a straightforward manner if open boundaries are chosen instead.
Generalizations to other dimensions and geometries are, in general, also straight-forward.

We label lattice sites as vectors of $d$ integers, $\site{x} \in \mathbb{Z}^d$, and use $\{ \hat{\site{e}}_k \}_{k=1 \dots d}$ as unit vectors in each positive direction. 
Links (edges between sites) are labeled as $\ell = (\site{x}, k)$. 
It is convenient to be able to refer to links from both of the adjacent sites; we therefore allow the index $k$ to range up to $2d$, and identify $\ell = (\site{x}, k)$ with the link in the $-\hat{\site{e}}_k$ direction from site $\site{x}$ for $k \in \{ 3,4 \}$ in two dimensions, as seen in figure \ref{fig:lattice_scheme}. In higher dimensions, some care must be taken when labelling the links, as we do when necessary (in~\cite{emonts_fermionic_2023}, the values $k \in \{ 5,6\}$ were used for the links in the $\pm \hat{\site{e}}_3$ directions).
When discussing a plaquette, we refer to the links as $p_i$, also shown in figure \ref{fig:lattice_scheme}. We also define $(-1)^{\site{x}} = (-1)^{\sum_k x_k}$ which takes the value $1$ on the even sublattice and $-1$ on the odd sublattice.

Our models of interest contain two types of degrees of freedom: fermionic matter residing on the sites and gauge fields residing on the links. 
As in the continuum formulation, the matter is represented by excitations in a fermionic Fock space; the gauge fields, however, as in the conventional compact lattice formulation~\cite{kogut_hamiltonian_1975}, are represented by first quantized Hilbert spaces on each link. 
In this section we will review all of these basic ingredients, first separately and then combined together into a lattice gauge theory, and introduce some notation and conventions used throughout the paper.

\subsection{The Gauge Group}
\label{sec:gauge-group}
We begin with the gauge group $G$ -- either a finite or a compact Lie group -- whose elements parameterize the symmetry transformations of the model. 
Its irreducible representations (irreps) will be denoted by the index (or collection of indices) $j$, and its elements by $g \in G$. Choices of different elements on each site will parameterize the local transformations.
A unitary representation $j$ of $g$ is given by the unitary matrix (sometimes called the \emph{Wigner matrix}) $D^{j}_{mn}\left(g\right)$, whose size is the dimension of $j$, denoted by $\text{dim}\left(j\right)$.

If $G$ is a compact Lie group, each $g \in G$ is uniquely determined by a set of \emph{group parameters}, or group coordinates, $\phi_a\left(g\right)$ (e.g. the Euler angles for $\mathit{SO}(3)$). 
Each Wigner matrix can then be expressed as
\begin{equation}
D^j(g)=e^{i\phi_a\left(g\right)T^j_a},
\end{equation}
where $T^j_a$ are the hermitian matrix representations of the group's generators, which satisfy the group's Lie algebra,
\begin{equation}
\left[T^j_a,T^j_b\right]=if_{abc}T^j_c,
\end{equation}
dictated by the group's structure constants $f_{abc}$ (e.g., for $\SU(2)$, $f_{abc}$ is the Levi-Civita symbol $\epsilon_{abc}$).

We will be particularly interested in the group $G=\SU(N)$ as well as $G=U(N)$. 
$\SU(N)$ has $N^2-1$ generators, $\left\{T_a\right\}_{a=1}^{N^2-1}$, which are all traceless, while $U(N)$ has an extra generator, $T_0 = \id$, the identity matrix of the appropriate dimension.
In particular, we will be interested in the fundamental representations of both groups, $j=N$, which are $N$ dimensional. 
They are accompanied by the conjugate representations, $\overline{N}$, which are also $N$ dimensional, and their generators are related to those of the fundamental irrep by 
\begin{equation}
    T^{j=\overline{N}}_a = -\overline{T}^{j=N}_a,
\end{equation}
where the overline on $T$ represents complex conjugation. 
In the $\SU(2)$ case, the fundamental representation is real, and thus these irreps are not independent.

\subsection{The Matter}
As stated above, the matter is fermionic, described in terms of a fermionic Fock space. 
On each site $\site{x}$ we introduce a set of Fock creation operators, $\psi^\dagger_{s m f}(\site{x})$, where the subscripts $s, m, f$ represent spin, color, and flavor respectively, which we take to be all of the intrinsic properties of the matter (additional properties could easily be added). Modes of different color interact differently with the gauge fields; modes with different flavor interact in the same way with gauge fields.

\subsubsection{Spins and Spacetime Symmetries}
The spin degree of freedom is, as usual, related to rotations. 
On a square lattice, the only possible rotations are around the main axes, by angles which are multiples of $\pi/2$. 
For $d=2$ we denote the $\pi/2$ rotation operator by $\mathcal{U}_R$, and we use it to implement the rotation,
\begin{equation}
    \mathcal{U}_R \
    \psi^\dagger_{s m f}(\site{x}) \ 
    \mathcal{U}^{\dagger}_R = \eta_{s s'}\left(\site{x}\right)\psi^\dagger_{s' m f}(\Lambda\site{x})
\end{equation}
where $\eta_{s s'}$ is a unitary rotation matrix mixing the spin indices and $\Lambda\site{x}$ is the coordinate rotation, $\Lambda\left(x_1,x_2\right)=\left(-x_2,x_1\right)$. 
In $d=3$ we have three non-commuting rotation operators, $\mathcal{U}^{(k)}_R$, as well as three non-commuting matrices 
$\eta^{(k)}_{s s'}(\site{x})$, such that
\begin{equation}
    \mathcal{U}^{(k)}_R \ 
    \psi^\dagger_{s m f}(\site{x}) \ 
    \mathcal{U}^{(k)\dagger}_R = \eta^{(k)}_{s s'}\left(\site{x}\right)\psi^\dagger_{s' m f}(\Lambda_k\site{x}),
\end{equation}
where $\Lambda_1\site{x} = \left(x_1,-x_3,x_2\right)$,
$\Lambda_2\site{x} = \left(x_3,x_2,-x_1\right)$, and
$\Lambda_3\site{x} = \left(-x_2,x_1,x_3\right)$.

It is well known that lattice fermions suffer from the doubling problem~\cite{susskind_lattice_1977,nielsen_no-go_1981}, in which unwanted spurious states appear on the lattice; various prescriptions are available for dealing with this issue. 
The framework introduced here is suitable for at least four prescriptions. 
The first two, with four spin components per site in $d=3$, are naive fermions (where nothing is done against doubling) and  Wilson's fermions~\cite{kogut_lattice_1983}. 
The other two involve staggering, where the spin components are distributed around several lattice sites (to be grouped together in the continuum limit). 
One can either use a staggering prescription with a single component per site~\cite{susskind_lattice_1977}, where the spin index becomes superfluous, or a partial staggering, with particle and anti-particles on two different sublattices~\cite{kogut_hamiltonian_1975}, giving rise to two-component spinors per site in $d=3$. 
A common property of all the prescriptions is that 
\begin{equation}
    [\eta^{(k)}]^4 = -\id,
\end{equation}
as expected from a $2\pi$ rotation of a fermion. 
Further details on rotations and  the $\eta^{(k)}$ matrices or factors, using the conventions of this work, may be found in~\cite{emonts_fermionic_2023}.

From now on, we shall focus on staggered fermions~\cite{susskind_lattice_1977} where there is no spin index at all (i.e. there is one spin component per link -- only color and flavor indices are left). 
We will also focus, for simplicity, on $d=2$.
Within these settings, $\eta$ is a phase, and we choose it, with no loss of generality, to be
\begin{equation}
    \eta\left(\site{x}\right)=\text{exp}\left( i\left(-1\right)^{\site{x}}\frac{\pi}{4}\right).
\end{equation}
For the generalizations to other formulations within the framework of PEPS, as well as to $d=3$, please refer to~\cite{emonts_fermionic_2023}. 
For our present purposes, which focus mostly on gauging, the staggered prescription is enough, as gauging involves the color indices alone, and in particular does not involve spin indices.

The choice of rotation prescription also dictates the form of translation invariance that we require. 
When the non-staggered prescriptions are used, we expect to have ``simple" translation invariance under single site translations. 
When staggering is involved  we expect invariance under two-site translations, since the unit cells are larger~\cite{susskind_lattice_1977}. 

\subsubsection{Colors and the Group Transformations}

In terms of the color index $m$, the creation operators on a given site $\psi^\dagger_{m f}(\site{x})$ form a complete spinor of some irrep $j$ of $G$. 
Usually this is taken to be the fundamental representation of the group, but this is not a requirement. 
This implies that $m$ varies over $\text{dim}\left(j\right)$ values, and that under group transformations parameterized by group elements $g\in G$ the creation operators get mixed. 
Quantitatively, for every $g\in G$ we define a unitary operator $\theta_g\left(\site{x}\right)$, such that
\begin{equation}
\theta_g\left(\site{x}\right)
\psi^\dagger_{m f}(\site{x})
\theta_g^{\dagger}\left(\site{x}\right)
=
\psi^\dagger_{n f}(\site{x})D^{j}_{nm}\left(g\right).
\end{equation}
The $\theta_g$ operators satisfy the group's  multiplication rule,
\begin{equation}
\theta_g\left(\site{x}\right)\theta_h\left(\site{x}\right)=\theta_{gh}\left(\site{x}\right).
\end{equation}
In the case of staggering~\cite{susskind_lattice_1977}, one needs to multiply this operator, on the odd sublattice, by  $\text{det}\left(D^j\left(g^{-1}\right)\right)$~\cite{zohar_formulation_2015}, which does not change the transformation rules of operators. 
This is related to the fact that on these sites, the absence of a fermion corresponds to the presence of an anti-particle.

One can generally include more than one irrep in a theory, as long as as each multiplet is taken as a whole. 
We will focus from now on (unless specified otherwise) on the case of $G$ which is either $\SU(N)$ or $U(N)$. 
We also assume that the matter is in the fundamental representation, with $N$ components. 
These cases are the most interesting for us physically, but generalizations to finite groups as well as other irreps are possible.

In the case of a compact Lie group $G$, we can express the $\theta_g$ operators using the group parameters $\phi_a\left(g\right)$ and the generating charges $Q_a$, which are defined as
\begin{equation}
    Q_a \left(\site{x}\right) = 
    \psi^{\dagger}_{ m f} \left(\site{x}\right)
    \left(T_a\right)_{mn} 
    \psi_{ n f} \left(\site{x}\right),
\end{equation}
and satisfy the Lie algebra
\begin{equation}
    \left[Q_a\left(\site{x}\right),Q_b\left(\site{y}\right)\right]=if_{abc}Q_c\left(\site{x}\right)\delta_{\site{x},\site{y} } 
\end{equation}
(no summation over $\site{x},\site{y}$).
If $G=\SU(N)$, all of the $N^2-1$ generators $\left\{T_a\right\}_{a=1}^{N^2-1}$ are traceless, and the representations of all group elements have determinant $1$. 
Thus, staggering does not affect the $\theta_g(\site{x})$ transformations, and can be expressed as
\begin{equation}
    \theta_g\left(\site{x}\right) = e^{i\phi_a\left(g\right)Q_a\left(\site{x}\right)}.
\end{equation}
For $G=U(N)$, this does not hold. 
Since $U(N)=\SU(N) \rtimes U(1)$, where the symbol $\rtimes$ denotes the semi-direct product, we can include the above transformations and the $N^2-1$ generators of $\SU(N)$, but we also need to include another generator, $T_0 = \id_N$, whose trace is $N$. 
Therefore we get, following~\cite{zohar_formulation_2015}, that for staggered $U(N)$,
\begin{equation}
    \theta_g\left(\site{x}\right) =  e^{i\phi_a\left(g\right)Q_a\left(\site{x}\right)}e^{-i\phi_0 Ns\left(\site{x}\right)},
\end{equation}
where $a \in \{0,...,N^2-1\}$ and $s\left(\site{x}\right)=0$ for even sites and $1$ for odd ones.

\subsubsection{Free Fermionic Theory}
Out of these ingredients we can now construct a free fermionic Hamiltonian, which will act on the matter alone, and which will later be gauged to include the fields on the links. 
It is
\begin{equation} \begin{aligned}
\label{eq:free-hamiltonian}
    H_0
    &= M\underset{\site{x}}{\sum}\left(-1\right)^{\site{x}}\psi_{mf}^{\dagger}\left(\site{x}\right)\psi_{mf}\left(\site{x}\right) 
    \\ &+ 
    \frac{i}{2a}\Bigg(
     \sum_{\site{x}} \bigg[ \psi_{mf}^{\dagger}\left(\site{x}\right)\psi_{mf}\left(\site{x}+\hat{\mathbf{e}}_1\right)
    \\ &+ i \left(-1\right)^{\site{x}}  \psi_{mf}^{\dagger}\left(\site{x}\right)\psi_{mf}\left(\site{x}+\hat{\mathbf{e}}_2\right) \bigg]
    -\text{h.c.}
    \Bigg),
\end{aligned} \end{equation}
where $a$ is the lattice spacing. The factors of $i$ are needed to ensure the Hamiltonian is invariant under rotations; this form matches that given in \cite{susskind_lattice_1977}, though we have generalized to include different possible values for color $m$ and flavor $f$. The version in \cite{susskind_lattice_1977} considers a 3D lattice; equation~\eqref{eq:free-hamiltonian} is restricted to the 2D case. In keeping with the convention throughout the paper, summation is implied over the repeated indices $m, f$. Here and below we have assigned equal masses to each flavor to simplify the presentation, but this can easily be generalized.

What are the symmetries of this Hamiltonian?
It is invariant under rotations as defined above, it is invariant under two site translations as expected in the staggered case, and it has a flavor permutation symmetry.
It is also invariant under charge conjugation, which is implemented by the single-site translations
\begin{equation}
    \psi_{mf}^{\dagger}\left(\site{x}\right) \rightarrow  
    \psi_{mf}\left(\site{x}+\hat{\mathbf{e}}_k\right),
    \quad k \in \{1,2\}.
\label{ccs}
\end{equation}

Another symmetry is the total fermion number conservation per flavor -- global $U(1)$ symmetries generated by
\begin{equation}
    N_{0,f}=\sum_{\site{x}} \psi^\dagger_{mf}(\site{x}) \psi_{mf}(\site{x}) 
\end{equation}
(no sum over $f$).
Because of this symmetry we focus on scenarios with a fixed fermionic filling, and often in particular on half-filling, which are dynamically connected to a Dirac-sea state (where all the even sites are empty and the odd ones are full)~\cite{susskind_lattice_1977} -- that is, states satisfying
\begin{equation}
N_0\left|\psi_0\right\rangle =\frac{\mathcal{N}_{\text{sites}}\mathcal{N}_{\text{flavors}}\mathcal{N}_{\text{colors}}}{2}\left|\psi_0\right\rangle,
\end{equation}
where $\mathcal{N}_{\text{sites}}$, $\mathcal{N}_{\text{flavors}}$, $\mathcal{N}_{\text{colors}}$ refer to the numbers of lattice sites, fermionic flavors, and colors respectively. 

We can add a chemical potential to the Hamiltonian, of the form 
\begin{equation} \label{eq:chemical-hamiltonian}
    H_{\mu} = \underset{\site{x}}{\sum}\mu_f\psi^{\dagger}_{mf}\left(\site{x}\right)\psi_{mf}\left(\site{x}\right),
\end{equation}
(now summing over both $m$ and $f$)
and it will be non-trivial only after coupling to a gauge field, since before doing so the Hamiltonian is just a sum of separate Hamiltonians of the different flavors. 
Only gauging, which we discuss below, will give rise to an indirect coupling of the different flavors (in the sense that they are all coupled to the same gauge field).

The Hamiltonian, with all terms included, is also invariant under global $G$ transformations (mixing the color indices),
\begin{equation}
    \theta_g = \underset{\site{x}}{\prod}\theta_g\left(\site{x}\right)
\end{equation}
for every $g \in G$. 
If $G=U(N)$, the $U(1)$ component of this symmetry is nothing but the total fermionic number conservation.

Being a free (quadratic) Hamiltonian, its ground state will be Gaussian, which can therefore be fully described by its covariance matrix (see appendix~\ref{sec:covariance}).

\subsection{The Gauge Field Hilbert Space}
Next we review the properties of the local gauge field Hilbert spaces, residing on the lattice's links, which will allow us to make the symmetry local (i.e. the system will be invariant under independent transformations at each site and surrounding links). 
On each link we introduce a Hilbert space which can be spanned by states labeled by elements of $G$ -- this gives the \emph{group element basis} $\{|g\rangle\}_{g\in G}$. 
Thus, for finite groups we have, on each link, a Hilbert space with a finite dimension which equals the group's order; in the case of infinite dimensional groups, the dimension will be infinite.

On each link, we can express the identity operator as
\begin{equation}
    \id = \int dg \ket{g}\bra{g},
    \label{groupcomp}
\end{equation}
where in the compact Lie case $dg$ is the Haar measure, and in the finite case the integral is replaced by a sum over all the group elements.
We have
\begin{equation}
    \left \langle g | h \right\rangle = \delta\left(g,h\right),
\end{equation}
where, if $G$ is finite $\delta\left(g,h\right)$ is the Kronecker delta, and otherwise it is a (Dirac delta) distribution defined with respect to the Haar measure.

We define \emph{gauge field configuration states} over the entire lattice as products of group element states on all the links, 
\begin{equation}
    \ket{\mathcal{G}}=\underset{\site{x},k}{\bigotimes}\ket{g\left(\site{x},k\right)}.
    \label{confdef}
\end{equation}
These states satisfy the  orthogonality condition
\begin{equation}
    \braket{\mathcal{G}}{\mathcal{G'}}=\underset{\site{x},k}{\prod} \delta\left(g\left(\site{x},k\right),g'\left(\site{x},k\right)\right)
\end{equation}
with respect to the appropriate $\delta$ as explained above, and where $g$, $g'$ are specified by $\mathcal{G}$, $\mathcal{G}'$.
We use $\int\mathcal{DG}$ to denote integration over all the gauge field configurations of the lattice (with the Haar measure) in the compact Lie case, and the corresponding summation in the finite group case.

On the Hilbert space of each link we define right and left group transformations, labelled by the group elements $g \in G$ and implemented by the unitary operators $\Theta_g$ and $\widetilde{\Theta}_g$ respectively. 
We can define them in terms of their action on the group element basis states,
\begin{equation}
    \Theta_g  |h\rangle = |h g^{-1}\rangle \qquad \qquad \widetilde\Theta_g  |h\rangle = |g^{-1}h\rangle.
    \label{Thetadef}
\end{equation}
These operators satisfy the group properties
\begin{equation}
    \begin{aligned}
        &\Theta_g \widetilde{\Theta}_h = \widetilde{\Theta}_h \Theta_g, \\
        &\Theta_g \Theta_h = \Theta_{gh}, \\
        & \widetilde{\Theta}_g  \widetilde{\Theta}_h =  \widetilde{\Theta}_{hg}.
    \end{aligned}
    \label{Thetarules}
\end{equation}

We introduce the connection or \emph{group element operator}, $U^j_{mn}$, which is a unitary matrix of link operators transforming under the irrep $j$:
\begin{equation}
    U^j_{mn} = \int dg \ D^j_{mn}(g)|g\rangle\langle g|.
    \label{eq:group_element_ops}
\end{equation}
While this is a matrix of operators, all of its elements commute, since they are all diagonal (as Hilbert space operators) in the group element basis. 
From equations~\eqref{Thetadef}, \eqref{Thetarules} and the fact that the $D$ matrices are unitary representations of $G$, it follows that
\begin{equation}
\begin{aligned}
    \Theta_g  U^j_{mn} \Theta^{\dagger}_g  &= U^j_{mn'} D^j_{n'n}\left(g\right), \\
    \widetilde{\Theta}_g  U^j_{mn} \widetilde{\Theta}^{\dagger}_g  &= D^j_{mm'}\left(g\right)U^j_{m'n} .
\end{aligned}
\end{equation}

Another useful basis is the conjugate one, usually referred to as the \emph{representation basis} or the irrep basis. 
Its elements are labelled by $\left|jmn\right\rangle$, where $j$
stands for an irrep, and $m$ and $n$ are multiplet indices corresponding to the eigenvalues of the maximal set of mutually commuting transformations for the left and right transformations respectively (e.g. the Cartan subalgebra in the case of compact Lie groups). 
The group transformations do not mix the $j$ multiplets,
\begin{equation} \begin{aligned} \label{eq:representation-basis}
    \Theta_g |jmn\rangle &= |jmn'\rangle D^j_{n'n}(g) \\
    \widetilde\Theta_g |jmn\rangle &= D^j_{mm'}(g) |jm'n\rangle.
\end{aligned} \end{equation}

Using the Peter-Weyl theorem, we can deduce the basis change formula
\begin{equation}
    \left\langle g | jmn\right\rangle =
    \sqrt{\frac{\text{dim}\left(j\right)}{\left|G\right|}} D^j_{mn}\left(g\right),
    \label{peterweyl}
\end{equation}
where $\left|G\right|$ is the group's order in the finite case, and the group's volume $\int dg$ (using the Haar measure $dg$) in the compact Lie case.

In the case of a compact Lie group, the operators $\Theta_g$ and $\widetilde\Theta_g$ can be expressed in terms of the group parameters and \emph{right} and \emph{left} generators, $R^a$ and $L^a$ respectively,
\begin{equation}
    \Theta_g=e^{i \phi_a(g) R^a}, \qquad \qquad \widetilde \Theta_g=e^{i \phi_a(g) L^a}.
\end{equation}
From the transformation properties we immediately get that
\begin{equation}
\begin{aligned}
    \left[R_a,R_b\right]&=i f^{abc}R_c, \\
    \left[L_a,L_b\right]&=-i f^{abc}L_c, \\
    \left[R_a,L_b\right]&=0,\\
    \left[L_a,U^j_{mn}\right]&=\left(T^j_a\right)_{mm'}U^j_{m'n},\\
    \left[R_a,U^j_{mn}\right]&=U^j_{mn'}\left(T^j_a\right)_{n'n}.    
\end{aligned}
\end{equation}

\begin{figure}[t]
    \centering
	\includegraphics[width=0.85\linewidth]{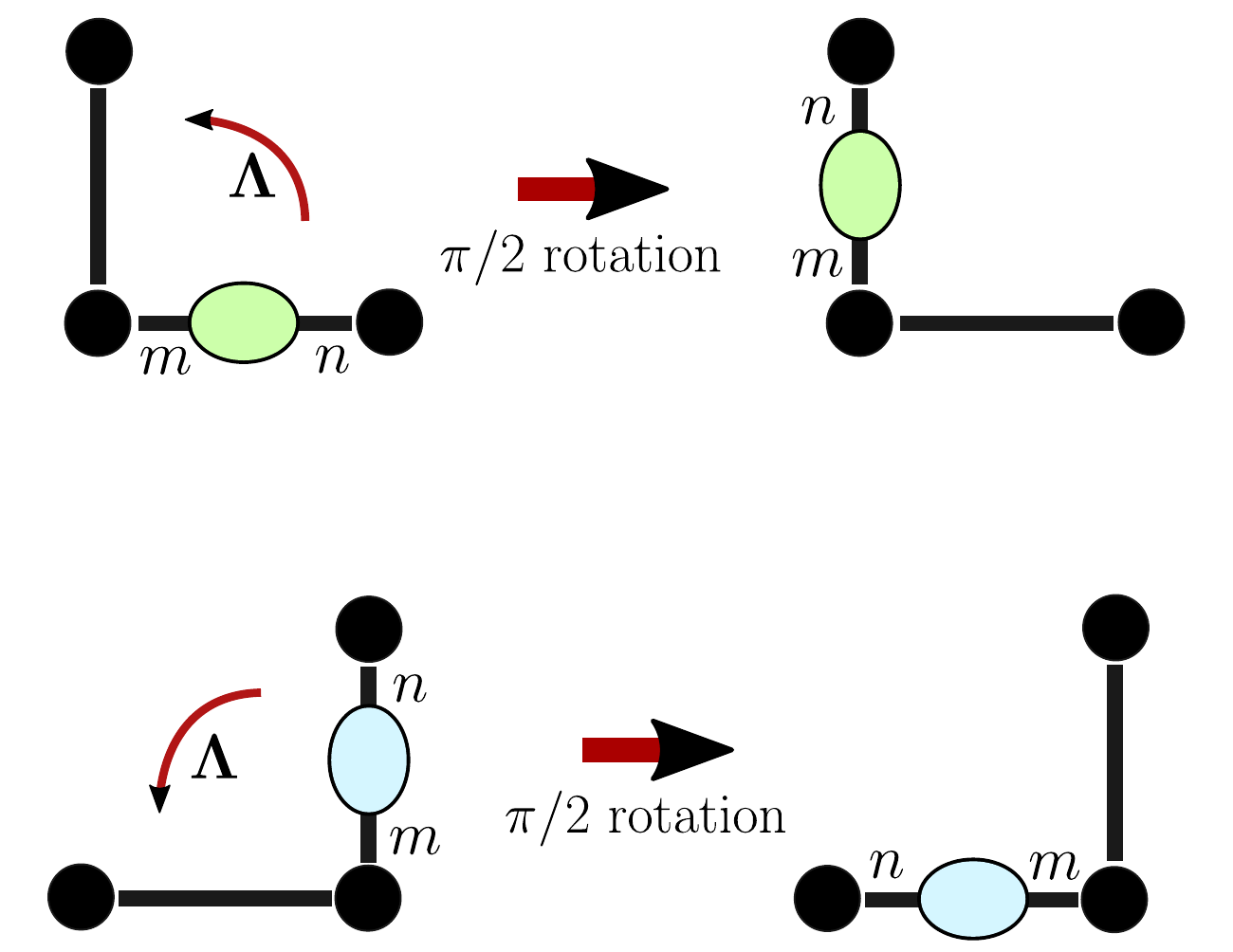}
	\caption{Schematic representation of a $\pi/2$ rotation $\mathbf{\Lambda}$ acting on the gauge field operators $U$ (colored ovals). The links are oriented left-to-right and bottom-to-top, and we assign the left and right quantum numbers $m$ and $n$ to the beginning and end of each link respectively. As shown, the rotation of the horizontal links is orientation-preserving, while for vertical links $m$ and $n$ are swapped. As a consequence, $U\rightarrow U^\dagger$ under this transformation.}
    \label{fig:rotations}
\end{figure}

As we shall review below, the group element operators will be useful in the Hamiltonian and other contexts within a lattice gauge theory, and thus we must address the issue of their rotation. 
We will see that the left quantum number, $m$, relates to the left/bottom side (or beginning) of the link, and the right quantum number, $n$, to the right/top side (the link's end). 
Since links (and gauge fields) are directional, we expect their rotation to have some sort of vector behavior. 
In 2D, for horizontal links,
\begin{equation}
    \mathcal{U}_R U^j_{mn}\left(\site{x},1\right) \mathcal{U}^{\dagger}_R = U^j_{mn}\left(\Lambda\site{x},2\right).
\end{equation}
Through this rotation, the beginning of the original link gets mapped to the beginning of the new one, and similarly the end -- this is an \emph{orientation preserving} rotation. 
Therefore we map $m$ to $m$ and $n$ to $n$.
However, the rotation of $U^j\left(\site{x},2\right)$ behaves differently. 
It can be seen graphically in figure~\ref{fig:rotations} that a vertical link will be rotated to a horizontal one, but with the opposite orientation -- that is, the beginning of the original link will be mapped to the end of the rotated one, and vice versa (compare with $\Lambda(x_1,x_2)=(-x_2,x_1)$). This means that the right and left Hilbert spaces of that link must be exchanged, and that right transformations have to be replaced by left ones. 
This is achieved by the transformation rule
\begin{align}
    \begin{aligned}
        \mathcal{U}_R U^j_{mn}\left(\site{x},2\right) \mathcal{U}^{\dagger}_R 
        &=  \overline{U}^j_{nm}\left(\Lambda\site{x}- \hat{\mathbf{e}}_1,1 \right)\\
        &= \left[\left(U^j\right)^{\dagger}\right]_{mn}\left(\Lambda\site{x}- \hat{\mathbf{e}}_1,1 \right).
    \end{aligned}
\label{Urotdef}
\end{align}
In $d=3$, rotations are defined in a similar way. 
For example, a rotation around the $z$ axis will behave like the $d=2$ version with the addition that $U^j_{mn}\left(\site{x},3\right) \rightarrow U^j_{mn}\left(\Lambda_3\site{x},3\right)$, and the other rotations are obtained by permutation.

\subsection{Minimal Coupling}
If we wish to lift the global $G$ symmetry of the free fermionic theory to a local one, we need to introduce connections, i.e. gauge fields, to our physical model. 
For this, we include the gauge field Hilbert spaces as introduced above, and modify the Hamiltonian to
\begin{equation} \begin{aligned}
	H&_\text{F} = M\underset{\site{x}}{\sum}\left(-1\right)^{\site{x}}\psi_{mf}^{\dagger}\left(\site{x}\right)\psi_{mf}\left(\site{x}\right)  \\&+
		\frac{i}{2a}\Bigg(
		 \sum_{\site{x}} \bigg[ \psi_{mf}^{\dagger}\left(\site{x}\right)
   U_{mn}\left(\site{x},1\right)
   \psi_{nf}\left(\site{x}+\hat{\mathbf{e}}_1\right)
		\\ &+
  i\left(-1\right)^{\site{x}} \psi_{mf}^{\dagger}\left(\site{x}\right)
        U_{mn}\left(\site{x},2\right)
    \psi_{nf}\left(\site{x}+\hat{\mathbf{e}}_2\right) \bigg]
		-\text{h.c.}
		\Bigg).
\end{aligned} \end{equation}
where the irrep index was omitted from the group element operators, which must be in the same irrep as the matter. 
We see, as expected, that $m$ is associated with the link's beginning and $n$ with its end.
The chemical potential term of equation~\eqref{eq:chemical-hamiltonian} is not modified, and we did not include it here for brevity.

This Hamiltonian remains translationally and rotationally invariant when the gauge field transformation rules are included. 
The fermionic global $U(1)$ symmetry is still present, and so is the flavor symmetry (if it is not broken by $H_{\mu}$). The global $G$ symmetry, however, has been lifted, as desired, to a local symmetry. 
We define the local, or gauge, transformations
\begin{equation}
    \hat{\Theta}_g(\site{x})=\prod_{k=1\dots d} \widetilde \Theta_g(\site{x},k) \Theta_g^\dagger(\site{x}-\hat{\mathbf{e}}_k,k)\theta^{\dagger}_g(\site{x}),
\end{equation}
and note that
\begin{equation}
    \left[H,\hat{\Theta}_g(\site{x})\right]=0,
\end{equation}
for every lattice site $\site{x}$ and $g \in G$.

This gives a local symmetry with conserved quantities on all the sites -- static charges, imposing a superselection rule on the Hilbert space. 
We focus, in general and unless specified otherwise, on the sector with no static charges, that is, the one with states $\ket{\Psi}$ satisfying
\begin{equation}
    \hat{\Theta}_g(\site{x})|\Psi\rangle = |\Psi\rangle
\end{equation}
for every lattice site $\site{x}$ and $g \in G$.
For gauge invariant operators $\mathcal{O}$,
\begin{equation}
    \hat{\Theta}_g(\site{x}) \mathcal{O}\,\hat{\Theta}^\dagger_g\,(\site{x})=\mathcal{O}
\end{equation}
is satisfied.

If $G$ is a compact Lie group, gauge invariance in the absence of static charges is equivalent to the \emph{Gauss laws},
\begin{equation}
    \begin{aligned}
        \left[\overset{d}{\underset{k=1}{\sum}}\big(
        L_a\left(\site{x},k\right)-
        R_a\left(\site{x}-\hat{\mathbf{e}}_k,k\right)\big)-
        Q_a\left(\site{x}\right)\right]
        \ket{\Psi}
        \equiv \\
        \big[\mathsf{D}_a\left(\site{x}\right) - Q_a\left(\site{x}\right)\big]\ket{\Psi}=0,\quad \forall \site{x},a,g\in G.
    \end{aligned}
    \label{Gausslaw}
\end{equation}
We can thus interpret the left and right generators as \emph{left and right electric fields}. We take this to define $\mathsf{D}_a(\site{x})$, the lattice divergence of the electric fields.

Since the only gauge field operators appearing in $H_\text{F}$ are group element operators $U$, $H_\text{F}$ clearly commutes with such operators, and therefore there is no gauge field dynamics. 
In order to include such dynamics, we add to $H_\text{F}$ the Kogut-Susskind Hamiltonian~\cite{kogut_hamiltonian_1975},
\begin{equation}
    H_{\text{KS}} = H_{\text{E}} + H_{\text{B}},
\end{equation}
which includes two terms, to which we give the monikers ``electric" and ``magnetic". These names comes from the interpretation of the terms in the case of a $U(1)$ gauge theory, where they correspond to the familiar electric and magnetic energies of electromagnetism.

The electric part is local, acting on each link separately, and takes the general form
\begin{equation}
 H_{\text{E}} = \lambda_{\text{E}}\underset{\site{x},k,j}{\sum} f_j \Pi_j\left(\site{x},k\right),
\end{equation}
where $k$ runs over the lattice directions, $j$ over the irreps, $\lambda_{\text{E}}$ and $f_j$ are real parameters, and $\Pi_j = \left|jmn\right\rangle\left\langle jmn \right|$ is a projector onto the $j$ irrep on a link. 
In the case of compact Lie groups, we use the quadratic Casimir operator $\mathbf{J}^2\equiv R_a R_a = L_aL_a$; e.g., for $\SU(2)$, $f_j=j\left(j+1\right)$.

The magnetic term involves four-body plaquette interactions, and takes (in $d=2$) the form
\begin{equation} \begin{aligned}
    H_{\text{B}} &= \lambda_{\text{B}}\sum_{\site{x}}
    \bigg( \text{Tr} \Big[ U\left(\site{x},1\right)U\left(\site{x}+\hat{\mathbf{e}}_1,2\right)
    \\ &\quad \times
    U^{\dagger}\left(\site{x}+\hat{\mathbf{e}}_2,1\right)U^{\dagger}\left(\site{x},2\right)\Big] + \text{h.c.} \bigg).
\end{aligned} \end{equation}
We have omitted the irrep index from the group elements again, assuming here (and below) that they belong to the same irrep as the matter.

The total Hamiltonian of a lattice gauge theory will thus take the form
\begin{equation}
    H=H_{\text{E}}+H_{\text{B}}+H_{\text{F}}+H_{\mu}.
    \label{Hbefore}
\end{equation}
It is still gauge invariant, as well as translation and rotation invariant. 
The fermionic symmetries are  unaffected by the inclusion of the pure gauge terms.

\section{Gauged Gaussian PEPS - Motivation} \label{sectionwhy}
With the lattice gauge theory background in mind, we now proceed to construct an ansatz state, to be used for variational ground state search. 
Generally, the ansatz state is embedded in the product space of the matter Fock spaces and the gauge field Hilbert spaces (within a particular static charge sector). 
Therefore, if we represent the identity using the completeness of the gauge field configuration basis,
\begin{equation}
    \id = \int \mathcal{DG}\ket{\mathcal{G}}\bra{\mathcal{G}},
    \label{compconf}
\end{equation}
we can generally expand our ansatz as
\begin{equation}
    \ket{\Psi} = \int \mathcal{DG} \ \ket{\mathcal{G}}\ket{\psi (\mathcal{G})},
    \label{Gexpansion_combined}
\end{equation}
where $\ket{\psi (\mathcal{G})}$ is the state of the matter fields.
As we discuss below, it will be convenient to include an explicit wavefunction of the gauge fields, giving
\begin{equation}
    \ket{\Psi} = \int \mathcal{DG} \ \psi_{I}\left(\mathcal{G}\right)\ket{\mathcal{G}}\ket{\psi_{II}\left(\mathcal{G}\right)},
    \label{Gexpansion}
\end{equation}
where $\psi_{I}\left(\mathcal{G}\right)$ is a wave function of the gauge fields in the configuration (group element) basis, and $\ket{\psi_{II}\left(\mathcal{G}\right)}$ is a state of the matter, describing fermions experiencing a static gauge field configuration $\mathcal{G}$. 
Neither is assumed to be normalized (and so $\ket{\Psi}$ is not either).
We now consider the requirements that we would like such a state to satisfy.

First, we would like it to capture relevant ground state physics (and hopefully that of other low lying states of the spectrum). 
Quantum information theory tells us that such states, when reasonable physical Hamiltonians are considered, are expected to satisfy the entanglement entropy area law \cite{hastings_area_2007,eisert_area_2010}, which is built into the construction of tensor network states, and in particular PEPS, projected entangled pair states~\cite{cirac_matrix_2021}. 
Therefore, we would like $\ket{\Psi}$ to be a PEPS.

Second, the state $\ket{\Psi}$ must be gauge invariant under $G$, and must exhibit all of the other symmetries discussed in the previous section. 
This is another good reason to use PEPS: they are suitable for encoding symmetries -- both global~\cite{cirac_matrix_2021} and local -- very naturally, through the use of gauging mechanisms of globally invariant PEPS~\cite{haegeman_gauging_2015,zohar_building_2016}. 
We will use the latter gauging method, which uses a fundamental analogy between PEPS and lattice gauge theories, as will be explained below. 

Third, the state $\ket{\Psi}$ should allow us to perform variational computations efficiently. 
We would like to be able to efficiently compute expectation values of gauge invariant operators, and in particular those appearing in the Hamiltonian, to be able to minimize the energy, find an approximate ground state, and study its physics. 
The relevant gauge invariant observables belong to three main classes: traces of oriented products of group element operators along closed paths, including the plaquette terms of the Hamiltonian, Wilson loops~\cite{wilson_confinement_1974}, and Polyakov loops (loops that wind around the periodic boundaries, and are therefore non-contractible); local terms diagonal in the representation basis, or electric field terms; and \enquote{mesonic operators} with oriented products of group element operators along an open path enclosed by fermionic operators. 
Let us examine the computation of the expectation value of all three with respect to our ansatz state $\ket{\Psi}$.

\begin{figure}[t]
\includegraphics[width=0.8\linewidth]  {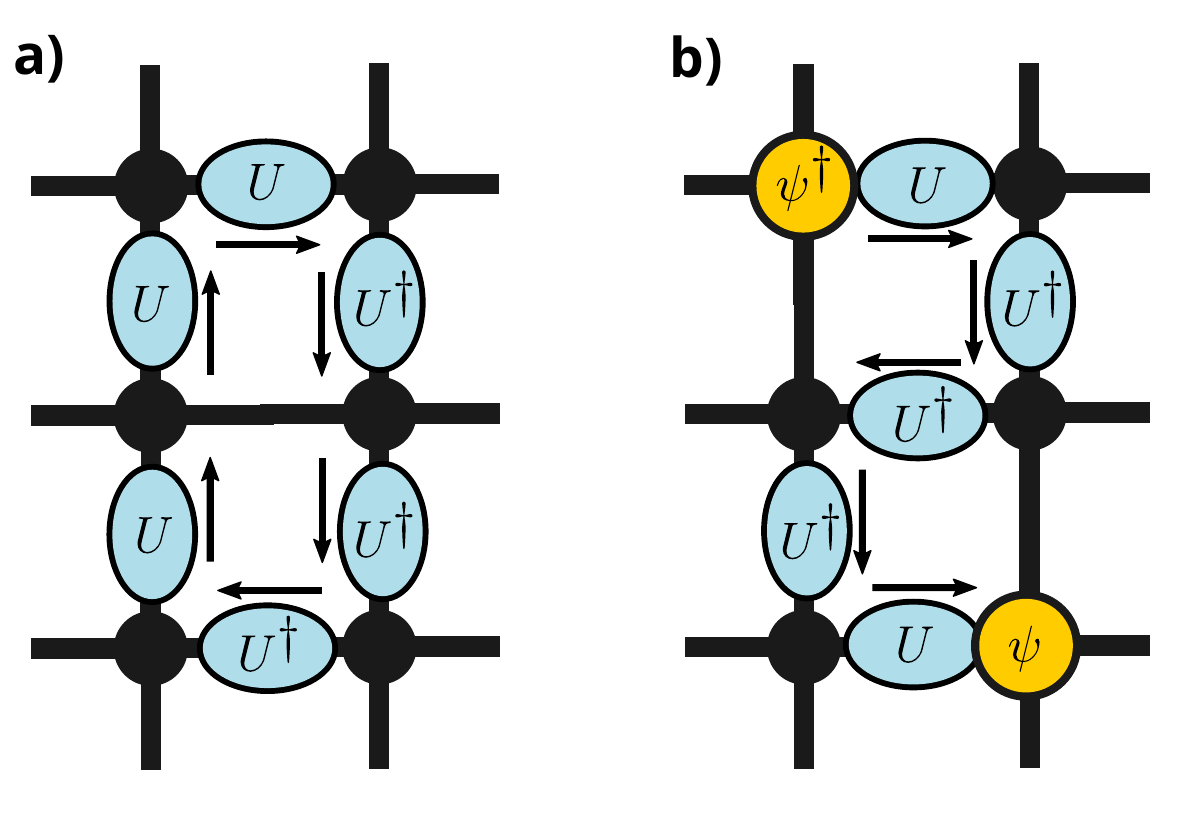}
\captionsetup[subfigure]{labelformat=empty}
    \subfloat[\label{subfig:observables_a}]{}
    \subfloat[\label{subfig:observables_b}]{}
	\caption{Examples of loop (left) and mesonic (right) operators that can be computed using the GGPEPS ansatz. Both closed and open paths are oriented, as shown by the arrows. The group element operator $U$ is taken when the path follows a link in the positive  direction (up or to the right), while in the remaining cases its conjugate $U^\dagger$ is taken.}
    \label{fig:observables}
\end{figure}

We begin with loop operators. 
Let $\mathcal{C}$ be some closed oriented path along the lattice. 
We define the loop operator along $\mathcal{C}$ as
\begin{equation}
W(\mathcal{C}) = \text{Tr}\left[\mathcal{P}\underset{\ell \in \mathcal{C}}{\prod}U\left(\ell\right)\right],
\label{WCdef}
\end{equation}
where we consider a matrix product of group element operators, involving their matrix indices. 
$\mathcal{P}$ stands for path ordering, modifying the terms: $U$ operators are taken on links in the positive direction ($k \in \{1,2 \}$) and $U^{\dagger}$ along links in the negative direction ($k \in \{3,4 \}$), as seen in Fig.~\ref{subfig:observables_a}.

The gauge field configuration states $\ket{\mathcal{G}}$ are eigenstates of these operators, satisfying
\begin{equation}
W(\mathcal{C})  \ket{\mathcal{G}} = \text{Tr}\left[\mathcal{P}\underset{\ell \in \mathcal{C}}{\prod}D\left(g_\ell\right)\right] \ket{\mathcal{G}}   ,
\end{equation}
where we now consider an oriented product of $D$ matrices. Therefore, using the orthogonality of the gauge field configuration states,
the expectation value of such a loop operator can be written as
\begin{equation} \begin{aligned}
    \left\langle W(\mathcal{C})  \right\rangle 
    &=
    \frac{\bra{\Psi}W\left(\mathcal{C}\right) \ket{\Psi}}{\braket{\Psi}{\Psi}} \\
    &=
    \frac{\int\mathcal{DG} \ \text{Tr}\left[\mathcal{P}\underset{\ell \in \mathcal{C}}{\prod}D\left(g_\ell\right)\right]\left|\psi_{I}\left(\mathcal{G}\right)\right|^2\braket{\psi_{II}\left(\mathcal{G}\right)}{\psi_{II}\left(\mathcal{G}\right)}}{\int\mathcal{DG'}\left|\psi_{I}\left(\mathcal{G'}\right)\right|^2\braket{\psi_{II}\left(\mathcal{G'}\right)}{\psi_{II}\left(\mathcal{G'}\right)}} 
    \\ &
    \equiv 
    \int\mathcal{DG} \ \text{Tr}\left[\mathcal{P}\underset{\ell \in \mathcal{C}}{\prod}D\left(g_\ell\right)\right] p\left(\mathcal{G}\right),
\end{aligned} \end{equation}
\label{FWdef}
where we have defined
\begin{equation}
 p\left(\mathcal{G}\right)=\frac{\left|\psi_{I}\left(\mathcal{G}\right)\right|^2\braket{\psi_{II}\left(\mathcal{G}\right)}{\psi_{II}\left(\mathcal{G}\right)}}{\int\mathcal{DG}'\left|\psi_{I}\left(\mathcal{G}'\right)\right|^2\braket{\psi_{II}\left(\mathcal{G}'\right)}{\psi_{II}\left(\mathcal{G}'\right)}}
 \equiv
 \frac{ p_0\left(\mathcal{G}\right)}{\mathcal{Z}}.
 \label{pdef}
\end{equation}
Note that $ p\left(\mathcal{G}\right)$ is a valid probability distribution over the gauge field configurations -- it is real, positive definite, and properly normalized since
$\int \mathcal{DG} \ p\left(\mathcal{G}\right)=1 $. 
Therefore, for a given $\ket{\Psi}$, 
$\left\langle W\left(\mathcal{C}\right)  \right\rangle$ may be evaluated using Monte-Carlo sampling, which will be \emph{sign-problem-free}.
Below, we will refer to $p_0\left(\mathcal{G}\right)$ as the \emph{unnormalized probability density}, and to $\mathcal{Z}$ as the \emph{partition function}.

The only possible bottleneck is in the computation of the norms defining $ p\left(\mathcal{G}\right)$: $\left|\psi_{I}\left(\mathcal{G}\right)\right|^2$ and $\braket{\psi_{II}\left(\mathcal{G}\right)}{\psi_{II}\left(\mathcal{G}\right)}$. 
This can be done by choosing $\ket{\Psi}$ to be a gauged Gaussian PEPS~\cite{zohar_fermionic_2015,zohar_projected_2016,zohar_combining_2018}. 
Then, as we shall review below, both norms are obtained from the contraction of Gaussian states, which is very efficient using the Gaussian formalism~\cite{bravyi_lagrangian_2005}.

The procedure, which will be described in the next subsections, involves the gauging of a Gaussian fermionic state representing the matter alone with a global symmetry, which is analogous to the procedure carried out when a free fermionic Hamiltonian is gauged -- another point in favor of this procedure.

The other observables may be computed very efficiently too, using a similar method. Consider, for example, the mesonic operator,
\begin{equation}
    \mathcal{M}_f\left(\site{x},\mathcal{C},\site{y}\right) = 
    \psi^{\dagger}_{mf}\left(\site{x}\right)
    \left[\mathcal{P}\underset{\ell \in \mathcal{C}}{\prod}U\left(\ell\right)\right]_{mn}
    \psi _{nf}\left(\site{y}\right),    
    \label{mesdef}
\end{equation}
where $\site{x},\site{y}$ are two lattice sites, $\mathcal{C}$ is some oriented path from $\site{x}$ to $\site{y}$, as seen in Fig.~\ref{subfig:observables_b}, and there is no summation over $f$.
We get, similarly, that
\begin{widetext}
\begin{equation}
\begin{aligned}
    \left\langle \mathcal{M}_f\left(\site{x},\mathcal{C},\site{y}\right) \right\rangle 
    &=
    \frac{1}{ \braket{\Psi}{\Psi} } \int \mathcal{DG} \left[\mathcal{P}\underset{\ell \in \mathcal{C}}{\prod}D\left(g_\ell\right)\right]_{mn}
    \bra{\psi_{II}\left(\mathcal{G}\right)} \psi^{\dagger}_{mf}\left(\site{x}\right)
        \psi _{nf}\left(\site{y}\right) \ket{\psi_{II}\left(\mathcal{G}\right)}\left|\psi_{I}\left(\mathcal{G}\right)\right|^2  \\
        &=
        \int \mathcal{DG} \left[\mathcal{P}\underset{\ell \in \mathcal{C}}{\prod}D\left(g_\ell\right)\right]_{mn}
    \frac{\bra{\psi_{II}\left(\mathcal{G}\right)} \psi^{\dagger}_{mf}\left(\site{x}\right)
        \psi _{nf}\left(\site{y}\right) \ket{\psi_{II}\left(\mathcal{G}\right)}}
        {\braket{\psi_{II}\left(\mathcal{G}\right)}} p\left(\mathcal{G}\right) 
        ,
        \label{meseval}
\end{aligned}
\end{equation}
\end{widetext}
with (again) no summation on $f$. 
This can also be computed using Monte-Carlo, and  the gauged Gaussian choice makes the computation of the new ingredient 
\begin{equation}
\frac{\bra{\psi_{II}\left(\mathcal{G}\right)} \psi^{\dagger}_{mf}\left(\site{x}\right)
        \psi _{nf}\left(\site{y}\right) \ket{\psi_{II}\left(\mathcal{G}\right)}}
        {\braket{\psi_{II}\left(\mathcal{G}\right)}}
\end{equation}
simple and efficient using elements of the covariance matrix (see appendix~\ref{sec:covariance}).

Finally, consider the electric field operators, or more generally functions thereof, $\mathcal{O}_{\text{E}}$. We have~\cite{zohar_combining_2018}
\begin{equation}
\left\langle \mathcal{O}_{\text{E}} \right\rangle =
\int \mathcal{DG} \ F_{\text{E}}\left(\mathcal{G}\right) p\left(\mathcal{G}\right),
\end{equation}
where 
\begin{equation}
F_{\text{E}}\left(\mathcal{G}\right) =\int \mathcal{D}\widetilde{\mathcal{G}} 
\bra{\widetilde{\mathcal{G}}}\mathcal{O}_{\text{E}}\ket{\mathcal{G}}
\frac{\overline{\psi_{I}}\left(\widetilde{\mathcal{G}}\right)\psi_{I}\left(\mathcal{G}\right)
\braket{\psi_{II}\left(\widetilde{\mathcal{G}}\right)}{\psi_{II}\left(\mathcal{G}\right)}}
{\left|\psi_{I}\left(\mathcal{G}\right)\right|^2\braket{\psi_{II}\left(\mathcal{G}\right)}{\psi_{II}\left(\mathcal{G}\right)}}.
\label{eintdef}
\end{equation}
This expectation value, involving overlaps of Gaussian states and wave functions in the gauged Gaussian case, can also be computed efficiently using Monte Carlo. 
Note that $\bra{\tilde{\mathcal{G}}}\mathcal{O}_{\text{E}}\ket{\mathcal{G}}$ can be computed analytically, and that generally the operators $\mathcal{O}_{\text{E}}$ are local, involving very few links (typically one). Therefore the integral in $F_{\mathcal{O}}\left(\mathcal{G}\right)$ can be done simply, as explained in~\cite{zohar_combining_2018} and demonstrated in~\cite{emonts_variational_2020,emonts_finding_2023}.

The expectation values of products of such operators are also easy to obtain in a similar fashion. In the case of more fermionic operators, Wick's theorem may be used~\cite{bravyi_lagrangian_2005}.

\section{Constructing a Globally Invariant Gaussian PEPS}

We now construct a state of the form of equation~\eqref{Gexpansion} as a PEPS.
Following references~\cite{zohar_fermionic_2015,zohar_projected_2016,zohar_combining_2018}, we start by constructing a state $\ket{\psi_0}$, representing the matter alone. It will be Gaussian -- a free state -- and invariant under all the global symmetries. Since the Hamiltonians of interest do not involve flavor mixing, we will begin by neglecting the flavor index and focusing on the color alone. 

As hinted in equation~\eqref{Gexpansion}, the ansatz state $\ket{\Psi}$ will be built in a way that will guarantee its decomposition to a product of two ingredients, for a fixed gauge field configuration $\mathcal{G}$ -- a pure-gauge wave function $\psi_{I}\left(\mathcal{G}\right)$ and a matter state $\ket{\psi_{II}\left(\mathcal{G}\right)}$. Following the standard technique to build Gaussian fermionic PEPS~\cite{kraus_fermionic_2010}, they will be constructed out of local Gaussian ingredients, involving both physical fermions and virtual modes.

On each site, we build a local state out of physical fermions together with  virtual modes (fermionic or bosonic, as discussed below) for each link adjacent to the site. 
As we explain, considerations relating to symmetries and expressive power of the ansatz motivate introducing multiple virtual modes per link.
As a first step we write down a free fermionic Gaussian PEPS with the desired global symmetries, which put constraints on the parametrization. 
The state is then gauged, which couples the state of the gauge field on each link with the virtual modes, guaranteeing that the Gauss Law is obeyed at each lattice site. 
Finally, the virtual modes are projected onto a maximally entangled state coupling the virtual modes from neighboring sites, and are then traced out, leaving a final physical state. In the following construction, we change the order of presentation, so that the gauging is discussed last. Before starting this construction, we introduce a convenient transformation that simplifies the construction and numerical calculations -- the particle-hole transformation.

\subsection{The Particle-Hole Transformation} \label{sec:ph-transformation}
The Gaussian PEPS we construct, introduced in~\cite{kraus_fermionic_2010}, has a BCS (Bardeen-Cooper-Schrieffer) form; that is, if we denote the Fock vacuum of the physical fermions introduced above by $\ket{\Omega_\text{p}}$, it will be of the form 
\begin{equation} \label{eq:bcs-form}
\exp\bigg(\underset{\site{x},\site{y}}{\sum}\mathcal{M}_{mn}\left(\site{x},\site{y}\right)
\psi^{\dagger}_{m}\left(\site{x}\right) \psi^{\dagger}_{n}\left(\site{y}\right)
 \bigg)\ket{\Omega_\text{p}}. 
\end{equation} 
Such states are not eigenstates of the total fermionic number operators, and thus cannot exhibit the $U(1)$ global symmetry introduced above. However, we can restore this symmetry by performing a \emph{particle-hole} transformation on the odd sublattice,
\begin{equation} \begin{aligned}
\label{parthole}
    \psi^{\dagger}_{ m}\left(\site{x}\right)
        &\to 
            \begin{cases}
            \psi^{\dagger}_{ m}\left(\site{x}\right) & \quad \site{x} \ \text{even} \\
            \psi_{ m}\left(\site{x}\right) & \quad \site{x} \ \text{odd}.
            \end{cases}
\end{aligned} \end{equation}
Before the transformation, the creation operators on the odd sublattice created holes: their absence corresponded to the presence of an anti-particle. 
In the new setting, they create anti-particles.

As a result, we get a few changes in the free fermionic theory discussed above. 
The free fermionic Hamiltonian will take a BCS form; in our staggered $d=2$ case, we will have (up to a constant)
\begin{equation} \begin{aligned}
\label{eq:free-hamiltonian-ph}
    H_0 &= M\underset{\site{x}}{\sum}\psi_{m}^{\dagger}\left(\site{x}\right)\psi_{m}\left(\site{x}\right) 
    \\ &+ 
		\frac{i}{2a}\bigg(
		 \underset{\site{x}}{\sum} \psi_{m}^{\dagger}\left(\site{x}\right)\psi^{\dagger}_{m}\left(\site{x}+\hat{\mathbf{e}}_1\right)
		\\ & +
  i \underset{\site{x}}{\sum} \psi_{m}^{\dagger}\left(\site{x}\right)\psi^{\dagger}_{m}\left(\site{x}+\hat{\mathbf{e}}_2\right)
		-\text{h.c.}
		\bigg).
\end{aligned} \end{equation}

What happens to the symmetries of the Hamiltonian? 
Note that due to the particle-hole transformation, the rotation phases on the odd sublattice must be conjugated, and given the choice for $\eta$ above, we now have a uniform, site-independent rotation phase of
\begin{equation}
    \eta\left(\site{x}\right) = e^{i\pi/4}.
\end{equation}
The new $H_0$ is invariant under a rotation with this phase.

One can also see that the two-site translation invariance has survived the particle-hole transformation; however,  the charge conjugation symmetry that we started with, which involved a particle-hole transformation, is now replaced by a simple, single-site translation invariance. 
That is, equation~\eqref{ccs} is replaced by
\begin{equation}
\psi_{m}^{\dagger}\left(\site{x}\right) \rightarrow  
\psi_{m}^{\dagger}\left(\site{x}+\hat{\mathbf{e}}_k\right),\quad k\in \{1,2\}.
\end{equation}

Next, we consider the global $G$ symmetry.
It does not matter whether $G = \SU(N)$ or $G = U(N)$, since even if we are only eventually interested in gauging $\SU(N)$, the global symmetry group at the moment is $U(N)$ due to the global $U(1)$ symmetry. 
We therefore discuss $U(N)$.

The generators on the odd sublattice undergo the transformation as
\begin{equation} \begin{aligned}
Q_a  &= 
\psi^{\dagger}_{ m} 
\left(T_a\right)_{mn} 
\psi_{n} 
\\ & \rightarrow
\psi_{ m} 
\left(T_a\right)_{mn} 
\psi^{\dagger}_{ n} 
= -\psi^{\dagger}_{ m} 
\left(\overline{T}_a\right)_{mn} 
\psi_{ n} +
\text{Tr}\left[T_a\right].
\end{aligned} \end{equation}
For the $\SU(N)$ generators, since $\text{Tr}\left[T_a\right]=0$, we simply get that the fermions on the odd sites now undergo the $G$ transformation with respect to the conjugate representation $\overline{N}$, instead of $N$, as expected for an anti-particle. 
For the extra $U(1)$ generator, we also get an exact cancellation of the staggering factor due to the trace of the identity.

We introduce the \emph{left charge},
\begin{equation}
\widetilde{Q}_a\left(\site{x}\right)  =
\psi^{\dagger}_{ m } \left(\site{x}\right)
\left(\overline{T}_a\right)_{mn} 
\psi_{ n } \left(\site{x}\right)    
\end{equation}
generating
\begin{equation}
\widetilde{\theta}_g\left(\site{x}\right) = e^{i\phi_a\left(g\right)\widetilde{Q}_a\left(\site{x}\right)},
\end{equation}
which implements left transformations on the matter,
\begin{equation}
    \widetilde{\theta}_g \psi^{\dagger}_{ m }  \widetilde{\theta}^{\dagger}_g =
    D_{mn}(g) \psi^{\dagger}_{ n } .
\end{equation}
This is useful, since $\widetilde{\theta}^{\dagger}_g$ implements the conjugate right transformations,
\begin{equation}
     \widetilde{\theta}^{\dagger}_g \psi^{\dagger}_{ m } \widetilde{\theta}_g =
     \psi^{\dagger}_{ n } \overline{D_{nm}(g)}.
\end{equation}
With this result we can write the generators of the global $G$ transformations as
\begin{equation}
    \mathcal{Q}_a = \underset{\site{x}\in\text{even}}{\sum}Q_a\left(\site{x}\right) - 
    \underset{\site{x}\in\text{odd}}{\sum}\widetilde{Q}_a\left(\site{x}\right),
\end{equation}
and we have
\begin{equation}    
\underset{\site{x}\in\text{even}}{\prod}\theta_g\left(\site{x}\right)
\underset{\site{x}\in\text{odd}}{\prod}\widetilde{\theta}^{\dagger}_g \left(\site{x}\right)
\ket{\psi_0} 
= e^{i\phi_a\mathcal{Q}_a}\ket{\psi_0}=\ket{\psi_0},
\label{zeroinv}
\end{equation}
where $\ket{\psi_0}$ is the state of the matter, which we will construct as a PEPS.

The special case of the subscript $a=0$ gives us the global $U(1)$ symmetry, which used to correspond to the conservation of total number of fermions. 
Under the particle hole transformation, 
\begin{equation}
    N_0 = 
    \sum_{\site{x}} \psi^\dagger_{m}(\site{x}) \psi_{m}(\site{x})
    \rightarrow \sum_{\site{x}} \left(-1\right)^{\site{x}}\psi^\dagger_{m}(\site{x}) \psi_{m}(\site{x})
    = \mathcal{Q}_0,
\end{equation}
up to a constant which equals the number of fermionic modes on the odd sublattice, due to which the half-filled sector becomes the sector with
\begin{equation}
    \mathcal{Q}_0\ket{\psi_0} = 0.
\end{equation}
The changes to the Gauss law and $H_\text{F}$ under the particle-hole transformation are given below in equations~\eqref{newgauss} and~\eqref{eq:fermionic-hamiltonian-after-ph}.

\subsection{The Virtual Modes and the PEPS Construction} \label{sec:peps-construction-main}
In order to construct a PEPS in the conventional way~\cite{cirac_matrix_2021} we introduce auxiliary, or virtual, degrees of freedom, which are responsible for contracting local degrees of freedom to a global state. 
Thus, in addition to the physical modes created by $\psi^{\dagger}_m$ on each site, we introduce auxiliary of virtual fermionic modes created by
$a^{\dagger}_{k \mu m}(\site{x})$. The indices $k, m$ maintain their earlier meaning labelling link and color; the new index $\mu$ labels ``copies", allowing for the possibility of multiple virtual modes per link.
Thus the $k$ index runs from $1$ to $2d$ and labels the links emanating from the site $\site{x}$ or ending there; for $d=2$, $k \in \{1,2,3,4\}$ corresponds to the links pointing towards $\hat{\mathbf{e}}_1$,$\hat{\mathbf{e}}_2$,$-\hat{\mathbf{e}}_1$,$-\hat{\mathbf{e}}_2$ respectively (for $d=3$ we add $k=5,6$, corresponding to $\hat{\mathbf{e}}_3$,$-\hat{\mathbf{e}}_3$~\cite{emonts_fermionic_2023}). 
We use $m$ to label the color components of all of the auxiliary modes, taking the same values as those taken by the physical modes. If another fermionic prescription is used for the physical fermions and they carry spin indices, the virtual ones will carry them as well, as in~\cite{emonts_fermionic_2023}. 
Finally, as mentioned, if we wish to include several copies of virtual modes, which will eventually enable us to increase the number of variational parameters, we can enumerate them by $\mu$.
We subsume a possible flavor index on the virtual modes into $\mu$; in a multi-flavor setting, we may choose to couple only certain copies to physical modes of each flavor. There is no requirement that the number of flavors and copies be the same.

It will be convenient to be able to refer to the physical and virtual modes together; we do so by defining
\begin{equation}
    \Psi^\dagger_\alpha (\site{x}) 
    \in \{ \psi^\dagger_m (\site{x}) \} \cup \{ a^{\dagger}_{k \mu m}(\site{x}) \}
\label{eq:all_modes}
\end{equation}
where $\alpha$ packages all the required indices.

Before describing the construction of the state in detail, we present the final form of the (ungauged) PEPS:
\begin{equation}
\ket{\psi_0} = 
\bra{\Omega_\text{v}} \prod_{(\site{x}, k)}
w^{\dagger}\left(\site{x}, k \right)
\underset{\site{x}}{\prod}A\left(\site{x}\right)\ket{\Omega_{\text{p}}} \ket{\Omega_{\text{v}}}
\label{eq:PEPSdef}
\end{equation}
where both $A(\site{x})$ and $w^{\dagger}\left(\site{x}, k \right)$ are Gaussian operators, and $\ket{\Omega_{\text{v}}}, \ket{\Omega_{\text{p}}}$ are the Fock vacua of virtual and physical modes respectively. $A(\site{x})$ takes the form
\begin{equation} \begin{aligned} \label{eq:A-op-full}
	A(\site{x})
    = \exp \Big( \mathcal{T}^{\alpha \beta} (\site{x})\Psi^\dagger_\alpha(\site{x}) \Psi^\dagger_\beta(\site{x}) \Big)
\end{aligned} \end{equation}
where $\mathcal{T}^{\alpha \beta} (\site{x}) \in \mathbb{C}$. This operator couples all the physical and virtual modes on site $\site{x}$, provided such coupling meets the symmetry conditions we discuss below.
The projection operators $w^{\dagger}\left(\site{x}, k \right)$ are defined as
\begin{equation} \begin{aligned}
\label{eq:projector-def}
w(\site{x}, k) &= \exp
\Bigg(
\xi
X^{(k)}_{ij} 
\underset{\mu}{\sum} W^{(k, \mu)}
a^{\dagger}_{i\mu m}\left(\site{x}\right)
a^{\dagger}_{j\mu m}\left(\site{x}+\hat{\mathbf{e}}_k\right)
\Bigg)
\end{aligned} \end{equation} 
where
\begin{equation} \begin{aligned} \label{Xdef}
X^{(1)}_{ij} &=\delta_{i,1}\delta_{j,3}, \\
X^{(2)}_{ij} &=\delta_{i,2}\delta_{j,4},
\end{aligned} \end{equation}
ensure that only virtual modes associated with the same links can couple; in three dimensions, we also include $X^{(3)}_{ij}  = \delta_{i,5}\delta_{j,6}$. $\xi$ and $W^{(k, \mu)}$ are complex numbers.
In keeping with our conventions, the sums are also over the repeated indices $i,j$.

To encode the desired symmetries in the PEPS, we need to specify the transformation properties of the virtual fermions, and then specify the constraints on the values $\mathcal{T}$ and $W$ which ensure the PEPS is invariant under all of the symmetries (the symmetries do no impose constraints on $\xi$).
We begin with the global $U(1)$ symmetry which guarantees particle number conservation.
For any virtual modes that couple to the physical matter, we must, following~\cite{emonts_fermionic_2023}, extend the global $U(1)$ symmetry. 
We therefore impose
\begin{equation} \label{eq:full-u1}
	\begin{aligned}
	\Psi^{\dagger}_\alpha\left(\site{x}\right) &\rightarrow \exp \left(i \chi_\alpha \left(-1\right)^{\site{x}}\varphi \right)\Psi^{\dagger}_\alpha\left(\site{x}\right).
	\end{aligned}
\end{equation}
The factor $(-1)^{\site{x}}$ accounts for the particle-hole transformation on odd sites, which adds a $-1$ factor to the phase gained under the $U(1)$ transformation.

In order to preserve the $U(1)$ symmetry, the modes which couple to physical matter must pick up a phase opposite to the one gained by that matter.
We therefore have three possible transformations under the $U(1)$ global symmetry (with a transformation parameter $\varphi$):
\begin{enumerate}
\item For modes that are entirely uncharged under this symmetry, i.e. the virtual modes that do not couple (even indirectly) to the physical modes, the $U(1)$ transformation does nothing.
\item \label{en:group2} Physical modes, and some virtual ones, will pick up a phase $\varphi$ on even sites and $-\varphi$ on odd sites.
\item \label{en:group3} Some virtual modes (those which couple to modes of the previous type) will pick up the phase $-\varphi$ on even sites and $\varphi$ on odd sites.
\end{enumerate}

The difference between even and odd sites arises due to the particle-hole transformation, as discussed in section~\ref{sec:ph-transformation}, and shown in equation \eqref{eq:full-u1}.
Accordingly, we define $\chi_\alpha$ to be $0, 1, -1$ in these cases respectively. Only modes $\alpha, \beta$ satisfying 
\begin{equation} \label{eq:u1-condition}
	\chi_\alpha + \chi_\beta = 0
\end{equation}
can couple to each other in the construction of the state.
We may have virtual modes that couple to the virtual modes in group \ref{en:group3}; these must then belong to group \ref{en:group2}, and therefore cannot themselves couple to the physical modes.
We will denote modes of the first group by $b^\dagger$, and refer to them as type I virtual modes.
In \cite{emonts_fermionic_2023}, the virtual modes of groups \ref{en:group2} and \ref{en:group3} were denoted by $d^\dagger$ and $c^\dagger$ respectively, and we adopt that convention moving forward, referring to them as type II virtual modes. The $b^\dagger, c^\dagger, d^\dagger$ modes are each different copies, but since there may be multiple copies of each type, we leave the $\mu$ subscript on these modes too.
It is clear from equation~\eqref{eq:u1-condition} that virtual modes of types I and II cannot couple to each other.

The possibility of virtual modes that are entirely uncoupled to physical modes (shown by the $V^{\text{PG}}$ block in equation~\eqref{eq:T-mat-illustration} below) is what allows for the separation of virtual modes into types I and II, which  build $\psi_I(\mathcal{G})$ and $\ket{\psi_{II}(\mathcal{G})}$  as given in equation~\eqref{Gexpansion}. Since the type I modes (those that are uncharged under the $U(1)$ transformation) do not couple to physical matter, it is not necessary that they be fermions -- type I virtual modes can be fermionic or bosonic. In fact, it is possible that there be both type I fermions and bosons (though such modes could not couple to each other in the construction of the state).
While fermionic virtual modes necessarily leads to a truncation of the gauge field Hilbert space (though this issue can be reduced by using extra copies of the virtual modes), virtual bosons avoid this issue, and allow for untruncated gauge fields even with a single copy. We leave a detailed discussion to future work.

It is possible to choose the type I modes $b^\dagger$ to be of any representation, which we label in this context by $r$: $\{b^{r\dagger}_{k\mu m}\left(\site{x}\right)\}$.
The only constraint is that all the color components $m$ must be included. 
It makes sense to include the fundamental representation again, but additional irreps can be used. 
Since these modes are completely decoupled from the rest, we may also use different numbers of copies. 
These modes will generate only \emph{pure gauge} excitations, that is, closed flux loops. 
The only charges they see are with respect to the gauge group $G$.

The virtual fields of type I form only closed loops (i.e. they involve no physical sources), while those of type II can also create open strings between charges. 
This motivates calling type I modes ``transversal", since they are source-free, and calling the type II modes ``longitudinal", in accordance with common phrasing in electrodynamics.

Next, we consider rotations, and introduce the 2D permutation matrix
\begin{equation}
\mathcal{R}_0 = \left( {\begin{array}{cccc}
		0 & 1 & 0 & 0 \\
		0 & 0 & 1 & 0 \\
		0 & 0 & 0 & 1 \\
		1 & 0 & 0 & 0 \\
\end{array} } \right),
\label{Rmat}
\end{equation}
which acts on the $k$ index and rotates the virtual modes around a site (ordered right, up, left, down). 
This is responsible for the spatial rotation. 
Following~\cite{emonts_fermionic_2023}, we define the rotation of the virtual modes as
\begin{equation}
	\begin{aligned}
	\mathcal{U}_V a^{\dagger}_{i \mu m}\left(\site{x}\right)\mathcal{U}^{\dagger}_V = \eta_\mu (\mathcal{R}_0)_{ij} a^{\dagger}_{j \mu m}\left(\Lambda\site{x}\right),
	\label{eq:rotv-general}
	\end{aligned}
\end{equation}
where $\eta_\mu$ is a phase that ensures  virtual fermions pick up a minus sign upon a full rotation. As indicated by the subscript, we do not require that all modes pick up the same phase under rotation. 
While one can choose the same phase under rotations for all modes when working on a two-dimensional lattice, in three dimensions accounting for spin and the non-commutativity of rotations makes it easier to handle phases separately for different copies, as done in \cite{emonts_fermionic_2023} and as we do here.
We thus make the choice that for virtual fermions,
\begin{equation} \begin{aligned}
 \mathcal{U}_V b^{r\dagger}_{i\mu m}\left(\site{x}\right)\mathcal{U}^{\dagger}_V = \eta (\mathcal{R}_0)_{ij} b^{r\dagger}_{j\mu m}\left(\Lambda\site{x}\right),
 \\
	\mathcal{U}_Vc^{\dagger}_{i\mu m}\left(\site{x}\right)\mathcal{U}^{\dagger}_V = \overline{\eta}(\mathcal{R}_0)_{ij}c^{\dagger}_{j\mu m}\left(\Lambda\site{x}\right),
 \\
		\mathcal{U}_Vd^{\dagger}_{i\mu m}\left(\site{x}\right)\mathcal{U}^{\dagger}_V = \eta(\mathcal{R}_0)_{ij}d^{\dagger}_{j\mu m}\left(\Lambda\site{x}\right).
	\label{rotv}
\end{aligned} \end{equation}
where $\eta$ is the phase picked up by physical modes as discussed above. 
For virtual modes that are bosons (which must be of type I), the rotation is given by simple permutations without extra phases,
\begin{equation}
\mathcal{U}_V b^{r\dagger}_{i\mu m}\left(\site{x}\right)   \mathcal{U}^{\dagger}_V  =
(\mathcal{R}_0)_{ij} b^{r\dagger}_{j\mu m}\left(\Lambda\site{x}\right),
\end{equation}
i.e. $\eta_\mu = 1$ for these copies.
Note that these rotations do not depend on the sublattice of $\site{x}$.

On each site, as in~\cite{emonts_fermionic_2023}, we define the Gaussian operators $A(\site{x})$ given in equation \eqref{eq:A-op-full}, which couples the modes $\alpha, \beta$ through the matrix $\mathcal{T}(\site{x})$.
To ensure translation invariance, we require $\mathcal{T} (\site{x}) = \mathcal{T}$, and to ensure rotation invariance, we require 
\begin{equation} \label{eq:T-rot-condition}
    \mathcal{R^\top TR} = \mathcal{T}
\end{equation} 
where $\mathcal{R}$ is the matrix that implements the rotations for all the modes $\Psi_\alpha^\dagger$.
In 2D, this matrix is given by
\begin{equation} \label{eq:full-rot-mat}
    \mathcal{R} = \begin{pNiceArray}{c|ccc}[margin,] 
                \eta \id & & & \\ \hline
                & \eta_\mu R_0 & & \\
                & & \Ddots^{} & \\
                & & & \eta_{\mu'} R_0
            \end{pNiceArray} 
\end{equation}
where the size of identity block $\eta \id$ is determined by the number of physical modes per site, and there are $\abs{ \{ a^\dagger_{ m, \mu } \} }$ copies of the $\eta_\mu \mathcal{R}_0$ blocks along the diagonal. 
All of the empty blocks are zero; the lines separate rows/columns of physical from virtual modes.
In 3D, the matrix blocks $\mathcal{R}_0$ must be expanded to include an identity block for the virtual modes oriented along the direction of rotation, as in \cite{emonts_fermionic_2023}. Further, one must account explicitly for spin indices, and satisfy equation~\eqref{eq:T-rot-condition} for rotations along all axes. Full details can be found in \cite{emonts_fermionic_2023}.

In 2D, satisfying equations~\eqref{eq:T-rot-condition} and~\eqref{eq:full-rot-mat} for copies $\mu$ and $\nu$ which pick up opposite phases under rotation, such as the $c^\dagger$ and $d^\dagger$ modes of equation~\eqref{rotv},
requires that the blocks $\tau^{(\mu,\nu)}$ of $\mathcal{T}$ be circulant matrices, i.e. of the form
\begin{equation} \label{eq:T-submat}
	\tau^{(\mu,\nu)} = \left( {\begin{array}{cccc}
			z_1^{(\mu,\nu)} & z_2^{(\mu,\nu)} & z_3^{(\mu,\nu)} & z_4^{(\mu,\nu)} \\
			z_4^{(\mu,\nu)} & 	z_1^{(\mu,\nu)} & z_2^{(\mu,\nu)} & z_3^{(\mu,\nu)} \\
			 z_3^{(\mu,\nu)} & z_4^{(\mu,\nu)} & z_1^{(\mu,\nu)} & z_2^{(\mu,\nu)} \\
			z_2^{(\mu,\nu)} & z_3^{(\mu,\nu)} & z_4^{(\mu,\nu)}  & z_1^{(\mu,\nu)} \\
	\end{array} } \right),
\end{equation}
where $z_i^{(\mu,\nu)} \in \mathbb{C}$. In \cite{emonts_finding_2023}, a pure-gauge theory was considered, without physical modes $\psi^\dagger$ or virtual modes of type II ($c^\dagger$ or $d^\dagger$). There, only (fermionic) modes $b^\dagger$ were considered, and they were taken to all pick up the same phase upon rotation. In such a case, the $\mathcal{T}$ matrix must compensate for the phases picked up by rotation, and factors of $\pm 1, \pm i$ also appear in $\tau^{(\mu,\nu)}$. Note that it is also possible to subdivide the uncharged (under the global $U(1)$ symmetry) $b^\dagger$ modes into copies of two subtypes which rotate with opposite phases, in which case the block $\tau^{(\mu,\nu)}$ will be of the form given in equation~\eqref{eq:T-submat}. 

For fermionic virtual modes, we also require that $\mathcal{T}^{\alpha \beta} = - \mathcal{T}^{\beta \alpha }$, i.e. $\mathcal{T}$ must be antisymmetric to be consistent with the fermionic commutation relations; for bosonic modes, $\mathcal{T}$ must be symmetric, $\mathcal{T}^{\alpha \beta} = \mathcal{T}^{\beta \alpha }$.
We leave a detailed discussion of bosonic virtual modes to future work. The motivations for including the virtual modes that are uncoupled to physical matter are discussed in detail below.

Accounting for all of these constraints, $\mathcal{T}$ has the following block form:\begin{equation} \label{eq:T-mat-illustration}
    \mathcal{T} = \begin{pNiceArray}{c|c:cc}[margin, first-row, 
        code-before = \rectanglecolor{red!15}{1-1}{1-1} 
            \rectanglecolor{green!15}{1-3}{1-3} 
            \rectanglecolor{green!15}{3-1}{3-1} 
            \rectanglecolor{orange!15}{2-2}{2-2}
            \rectanglecolor{blue!15}{3-3}{4-4}] 
        \psi & b & c & d \\
        \mathcal{T}^{PP}         & \mat{0}               & M           & \mat{0}  \\ \hline
        \mat{0}        & V^{\text{PG}} & \mat{0}       & \mat{0}  \\ \hdottedline
        -M^\top        &  \mat{0}              & \mat{0}       & \tilde{V}   \\ 
        \mat{0}        & \mat{0}               & - \tilde{V}^\top  & \mat{0}  \\              
    \end{pNiceArray} 
\end{equation}
where the $\mathcal{T}^{PP}$ block above and to the left of the solid lines couples physical modes among themselves, the $V^\text{PG}$ block bounded by the solid and dotted lines couples virtual modes (fermionic or bosonic) that do not couple to physical fermions, and the bottom right block containing both copies of $\tilde{V}$ shows the virtual modes which do couple to physical fermions, as shown in the $M$ blocks above and to the left of the solid lines. If one wishes to allow for virtual-virtual coupling while maintaining the $U(1)$ symmetry, the $\tilde{V}$ block must contain (a minimum of) two copies, only one of which couples to the physical modes (as shown). Note that the constraints due to rotation invariance must be imposed on the submatrices.

Note that $[A(\site{x}), A(\site{y})] = 0$ for any two sites $\site{x},\site{y}\in\mathbb{Z}^d$, and therefore the product $\prod_{\site{x}} A(\site{x})$ is well defined and requires no ordering.
Since the type I and type II virtual modes do not couple, we can divide $A(\site{x})$ into $A^{(I)}(\site{x})$
and $A^{(II)}(\site{x})$ which are each of the form of equation~\eqref{eq:A-op-full}, but include only the modes of type I and II respectively. These operators satisfy
\begin{equation} \label{eq:A-op-combined}
    A\left(\site{x}\right) = A^{(I)}\left(\site{x}\right) A^{(II)}\left(\site{x}\right).
\end{equation}
Note that all of these operators commute.

If we account for the separation of the type I and II modes, as well as the constraints of the $U(1)$ symmetry (which account for the $\mat{0}$ blocks in $\mathcal{T}$), we can write
\begin{equation} \begin{aligned}
    A^{(I)}\left(\site{x}\right) = \exp \Bigg( & \sum_{r, \mu, \nu}
    \tau^{\left(\mu,\nu\right)}_{ij,mm'}
    b^{r\dagger}_{i\mu m}\left(\site{x}\right)
    b^{r\dagger}_{j\nu m'}\left(\site{x}\right)
    \Bigg),
\\
A^{(II)}\left(\site{x}\right) = \exp \Bigg( & \sum_{\mu}
t_i^{(\mu)} \psi^{\dagger}_{m}\left(\site{x}\right)c^{\dagger}_{i\mu m}\left(\site{x}\right) 
\\ +
& \sum_{\mu, \nu}
\tau^{(\mu,\nu)}_{ij}
c^{\dagger}_{i \mu m}\left(\site{x}\right)
d^{\dagger}_{j\nu m}\left(\site{x}\right)\Bigg),
\label{eq:A-def-types}
\end{aligned} \end{equation}
where the sums are also over the link directions $i, j$ and the colors $m, m'$. The $t_i^{(\mu)} \in \mathbb{C}$ are elements of the block $M$ shown in equation~\eqref{eq:T-mat-illustration}. The copies $\mu, \nu$ that are included in each operator are of course limited to those of the appropriate type.

The direct coupling of physical fermions to themselves is absent here (though it could easily be added), since it is not expected in this staggered formulation. Direct coupling of different flavors is not expected in this case, because the free fermionic Hamiltonian is a sum of separate Hamiltonians (given in equation~\eqref{eq:free-hamiltonian-ph} above) for each flavor, and the state is expected to be a product state of the flavors, as explained above. This continues to hold after gauging, as explained below.

Next, on each link define the Gaussian projection operators $w(\site{x}, k)$ given in equation~\eqref{eq:projector-def}.
The values of $W^{(k)}$ must be chosen in a way that guarantees that $w(\site{x}, k)$ is invariant under all the symmetries discussed above. 
We've also suppressed any dependence on the location of the link (through $\site{x}$) in order to preserve invariance under translations; the dependence of $W^{(k)}$ on the link direction is necessary to guarantee invariance under rotations.
$\xi$ can be any arbitrary constant; in~\cite{emonts_finding_2023}, $\xi = 1$ was used, while $\xi=i$ was used in~\cite{emonts_fermionic_2023}, both for the sake of convenience.

As with the operators $A(\site{x})$, type I and II virtual modes cannot couple in $w(\site{x}, k)$, and so we can divide them into $w^{(I)}(\site{x}, k)$ and $w^{(II)}(\site{x}, k)$, with 
\begin{equation}
    w\left(\site{x},k\right) =w^{(I)}\left(\site{x},k\right)w^{(II)}\left(\site{x},k\right).
\end{equation}
These operators project virtual modes from adjacent sites on the same link onto a maximally entangled state. This is what ensures that our state obeys an entanglement area law -- the entanglement between any two subsystems only arises due to this entanglement on links, and so the total entanglement is  proportional to the number of links between to subsystems, i.e. to the ``area" of the boundary between them.

Accounting for the constraints due to rotation and the $U(1)$ symmetry, these can be written as
\begin{equation} \begin{aligned}
w^{(I)}\left(\site{x}, k \right) &= \exp
\bigg(
\xi
\hat{W}^{(k)}X^{(k)}_{ij} 
\underset{r,\mu}{\sum}
b^{r\dagger}_{i\mu m}\left(\site{x}\right)
b^{r\dagger}_{j\mu m}\left(\site{x}+\hat{\mathbf{e}}_k\right)
\bigg)
\\
w^{(II)}\left(\site{x}, k \right) &= \exp
\bigg(
\xi
\bar{W}^{(k)}X^{(k)}_{ij} 
\underset{\mu}{\sum}
c^{\dagger}_{i\mu m}\left(\site{x}\right)
c^{\dagger}_{j\mu m}\left(\site{x}+\hat{\mathbf{e}}_k\right)
\bigg)
\\ \times
\exp &
\bigg(
\xi
\tilde{W}^{(k)}X^{(k)}_{ij} 
\underset{\mu}{\sum}
d^{\dagger}_{i\mu m}\left(\site{x}\right)
d^{\dagger}_{j\mu m}\left(\site{x}+\hat{\mathbf{e}}_k\right)
\bigg),
\label{wdef}
\end{aligned} \end{equation} 
where we have chosen $W^{(k, \mu)} = W^{(k)}$, and suppressed the $\mu$ index.
Choosing
\begin{equation}
\begin{aligned}
\hat{W}^{(1)} = 1, &\quad   \hat{W}^{(2)} = \eta^2, \\
\bar{W}^{(1)} = 1, &\quad   \bar{W}^{(2)} = \overline{\eta}^2, \\
\tilde{W}^{(1)} = 1, &\quad   \tilde{W}^{(2)} = \eta^2,
\end{aligned}
\end{equation}
for fermions, and 
\begin{equation}
\begin{aligned}
\hat{W}^{(1)} = \hat{W}^{(2)} = 1, 
\end{aligned}
\end{equation}
for bosons, 
guarantees the invariance of $w^{(I)}(\site{x}, k)$ and $w^{(II)}(\site{x}, k)$ (and therefore also $w(\site{x}, k)$) under rotations and the $U(1)$ symmetry. Other choices of $W^{(k, \mu)}$ are also possible.
Note that while the $c^\dagger$ and $d^\dagger$ modes could not couple to themselves in $A(\site{x})$, here they do -- and must -- since the modes belong to neighboring sites (one on the even, one on the odd, sublattice), and therefore pick up opposite phases upon rotation and under the $U(1)$ transformation.
All of these definitions and settings are from~\cite{emonts_fermionic_2023}, which can be used, as usual, to generalize the construction to $d=3$ and other spin prescriptions. 
Choosing rotations other than those in equation~\eqref{rotv}, would lead to different choices for $W^{(k, \mu)}$ (in addition to the need, mentioned above, to absorb the phases in $\mathcal{T})$.

All of the $w$ operators of different links mutually commute.
Therefore, no ordering of products is required for constructing the (ungauged) Gaussian fermionic PEPS, given in equation~\eqref{eq:PEPSdef}.
This state is a product of local physical Gaussian states created by the $A\left(\site{x}\right)$ on the physical and virtual Fock vacuum, contracted by virtual maximally-entangled states on the links created by the $w\left(\site{x}, k \right)$ operators.

Taking advantage of the separation of virtual modes into types I and II, the  state of the matter can be written as
\begin{equation} \begin{aligned}
\ket{\psi_0} &= \psi_{I} \ket{\psi_{II}} 
\end{aligned} \end{equation}
where
\begin{equation} \begin{aligned}
\psi_{I} = 
\bra{\Omega_{I}} \underset{\site{x},k}{\prod}w^{(I)\dagger}\left(\site{x}, k\right)
\underset{\site{x}}{\prod}A^{(I)}\left(\site{x}\right)\ket{\Omega_{I}}
\end{aligned} \end{equation}
and
\begin{equation} \begin{aligned}
\ket{\psi_{II}} = 
\bra{\Omega_{II}} \underset{\site{x},k}{\prod}w^{(II)\dagger}\left(\site{x}, k\right)
\underset{\site{x}}{\prod}A^{(II)}\left(\site{x}\right)\ket{\Omega_{\text{p}}}\ket{\Omega_{II}},
\end{aligned} \end{equation}
where $\ket{\Omega_I}$ and $\ket{\Omega_{II}}$ are the Fock vacua for the type I and II virtual modes respectively. 
Because the virtual modes are traced out, $\psi_I$ is simply a number, while $\ket{\psi_{II}}$ is a state of the physical fermions.

As a result, we get a state $\ket{\psi_0}$ containing the physical degrees of freedom alone, with an entanglement entropy area law -- a Gaussian fermionic PEPS~\cite{kraus_fermionic_2010}. 
Having followed the construction of~\cite{emonts_fermionic_2023}, we are guaranteed that this state is rotationally invariant and respects the global $U(1)$ symmetry. It is also translationally invariant and obeys fermionic statistics.

\subsection{Global $G$ Invariance}

The PEPS $\ket{\psi_0}$ is invariant under single site translations,
\begin{equation} \begin{aligned}
\psi_{m}^{\dagger}\left(\site{x}\right) &\rightarrow  
\psi_{m}^{\dagger}\left(\site{x}+\hat{\mathbf{e}}_k\right),\\
a^{\dagger}_{k\mu m}\left(\site{x}\right) &\rightarrow  
a^{\dagger}_{k\mu m}\left(\site{x}+\hat{\mathbf{e}}_k\right).
\end{aligned} \end{equation}
In the case of physical fermions, we already know that the particle-hole transformation corresponds to charge conjugation. 
We can therefore give the same interpretation to the virtual fermions, which fits well with the way they participate in the global $U(1)$ symmetry. 
Thus for the group transformations we require that the $c^\dagger$ fermions undergo the global $U(N)$ transformations, including the $\SU(N)$ ones, with respect to the representation conjugate to that of the physical ones, and the $d^\dagger$ fermions under the same representation as the physical ones, which is summarized in table~\ref{table:u1-transformation-modes}. This will ensure invariance of both $A(\site{x})$ and $w(\site{x}, k)$ under a global transformation parameterized by $g \in G$.
We must also specify how the $b^\dagger$ modes, which don't couple to physical modes transform under a group transformation, but there are no constraints that arise in that case from coupling to physical matter. 

\begin{table}[H] 
    \centering
    \vspace{5mm}
    \begin{tabular}{l||c|c}
    & even sublattice  & odd lattice
    \\ \hline \hline
    $\psi_{m}^{\dagger}\left(\site{x}\right)$      & $N$ (fundamental) & $\overline{N}$ (anti-fundamental) 
    \rule[-1ex]{0pt}{4ex} \\ \hline
    $c^{\dagger}_{k\mu m}\left(\site{x}\right)$   & $\overline{N}$ (anti-fundamental) & $N$   (fundamental) 
    \rule[-1ex]{0pt}{4ex} \\ \hline
    $d^{\dagger}_{k\mu m}\left(\site{x}\right)$ & $N$ (fundamental)  & $\overline{N}$ (anti-fundamental)
    \rule[-1ex]{0pt}{4ex} \\ \hline
    \end{tabular}
    \caption{Representations under which the physical and type II virtual fermions undergo $U(N)$ transformations, after the particle-hole transformation, on both sublattices. The $b^\dagger$ modes transform according to their irrep $r$.}
     \label{table:u1-transformation-modes}
\end{table}

In order to define these transformations, we introduce what will be their generators, which we call \emph{virtual electric fields},
\begin{equation}
\begin{aligned}
E_a^C\left(\site{x},k\right) &= 
\underset{\mu}{\sum}c^{\dagger}_{k\mu m} \left(\site{x}\right) \left(T_a\right)_{mn} 
c_{k\mu n} \left(\site{x}\right), \\
\widetilde{E}_a^C\left(\site{x},k\right) &= 
\underset{\mu}{\sum}c^{\dagger}_{k\mu m} \left(\site{x}\right) \left(\overline{T}_a\right)_{mn} 
c_{k\mu n} \left(\site{x}\right), \\
\end{aligned}
\label{ECDdef}
\end{equation}
with no summation over $k$, and where $T_a$ is defined in section~\ref{sec:gauge-group}. We also define $E_a^B(\site{x},k)$, $\widetilde{E}_a^B(\site{x},k)$, $E_a^D(\site{x},k)$, and $\widetilde{E}_a^D(\site{x},k)$ in the same way, replacing the modes $c^\dagger$ with $b^\dagger$ and $d^\dagger$ respectively.
They satisfy the right and left group algebras,
\begin{equation} \begin{aligned}
\left[E_a^C,E_b^C\right]&=if_{abc}E_c^C, \\
\left[\widetilde{E}_a^C,\widetilde{E}_b^C\right]&=-if_{abc}\widetilde{E}_c^C, \\
\label{ECDalg}
\end{aligned} \end{equation}
and similarly for the $b^\dagger$ and $d^\dagger$ modes.
Therefore, they generate right and left group transformations of these modes, similarly to those defined for the physical fermions.

Next, we define virtual electric fields, which will be converted to physical ones through the gauging procedure, 
\begin{equation} \begin{aligned} \label{E1def}
    E^{(I)}_a\left(\site{x},k\right)
    &= E_a^B\left(\site{x},k\right), \\
    \widetilde{E}^{(I)}_a\left(\site{x},k\right)
    &= \widetilde{E}_a^B\left(\site{x},k\right),
\end{aligned} \end{equation}
and
\begin{equation} \begin{aligned}
\label{E2def}
    E^{(II)}_a\left(\site{x},k\right)
    &= E_a^C\left(\site{x},k\right) - \widetilde{E}_a^D\left(\site{x},k\right), \\
    \widetilde{E}^{(II)}_a\left(\site{x},k\right)
    &= \widetilde{E}_a^C\left(\site{x},k\right) - E_a^D\left(\site{x},k\right),
\end{aligned} \end{equation}
and with these, we introduce the finite transformations
\begin{equation}
\begin{aligned}
        \theta^{(J)}_g\left(\site{x},k\right) &= e^{i\phi_a E^{(J)}_a\left(\site{x},k\right)}, \\
    \widetilde{\theta}^{(J)}_g\left(\site{x},k\right) &= e^{i\phi_a \widetilde{E}^{(J)}_a\left(\site{x},k\right)},
    \label{theta2def}
\end{aligned}
\end{equation}
where $J \in \{ I, II\}$.
These satisfy
\begin{equation} \begin{aligned}
\label{globinv}
\theta^{(I)}_g\left(\site{x},k\right) 
b^{r\dagger}_{k\mu m}\left(\site{x}\right)
\theta^{(I)\dagger}_g\left(\site{x},k\right) 
&=b^{r\dagger}_{k\mu n}\left(\site{x},k\right) D^r_{nm}(g),\\
\widetilde{\theta}^{(I)}_g\left(\site{x},k\right) 
b^{r\dagger}_{k\mu m}\left(\site{x}\right)
\widetilde{\theta}^{(I)\dagger}_g\left(\site{x},k\right) 
&=D^r_{mn}(g)b^{r\dagger}_{k\mu n}\left(\site{x},k\right) \\
\theta^{(II)}_g\left(\site{x},k\right) 
c^{\dagger}_{k\mu m}\left(\site{x}\right)
\theta^{(II)\dagger}_g\left(\site{x},k\right) 
&=c^{\dagger}_{k\mu n}\left(\site{x},k\right) D_{nm}(g),\\
\widetilde{\theta}^{(II)}_g\left(\site{x},k\right) 
c^{\dagger}_{k\mu m}\left(\site{x}\right)
\widetilde{\theta}^{(II)\dagger}_g\left(\site{x},k\right) 
&=D_{mn}(g)c^{\dagger}_{k\mu n}\left(\site{x},k\right) ,\\
\theta^{(II)}_g\left(\site{x},k\right) 
d^{\dagger}_{k\mu m}\left(\site{x}\right)
\theta^{(II)\dagger}_g\left(\site{x},k\right) 
&=d^{\dagger}_{k\mu n}\left(\site{x},k\right)\overline{D_{nm}(g)},\\
\widetilde{\theta}^{(II)}_g\left(\site{x},k\right) 
d^{\dagger}_{k\mu m}\left(\site{x}\right)
\widetilde{\theta}^{(II)\dagger}_g\left(\site{x},k\right) 
&= \overline{D_{mn}(g)}d^{\dagger}_{k\mu n}\left(\site{x},k\right) 
\end{aligned} \end{equation}
(no summation over $k$). Though this looks dense, it simply details how the virtual modes are transformed by a global transformation parameterized by group element $g$.

\begin{figure}[h]
\includegraphics[width=0.8\linewidth]{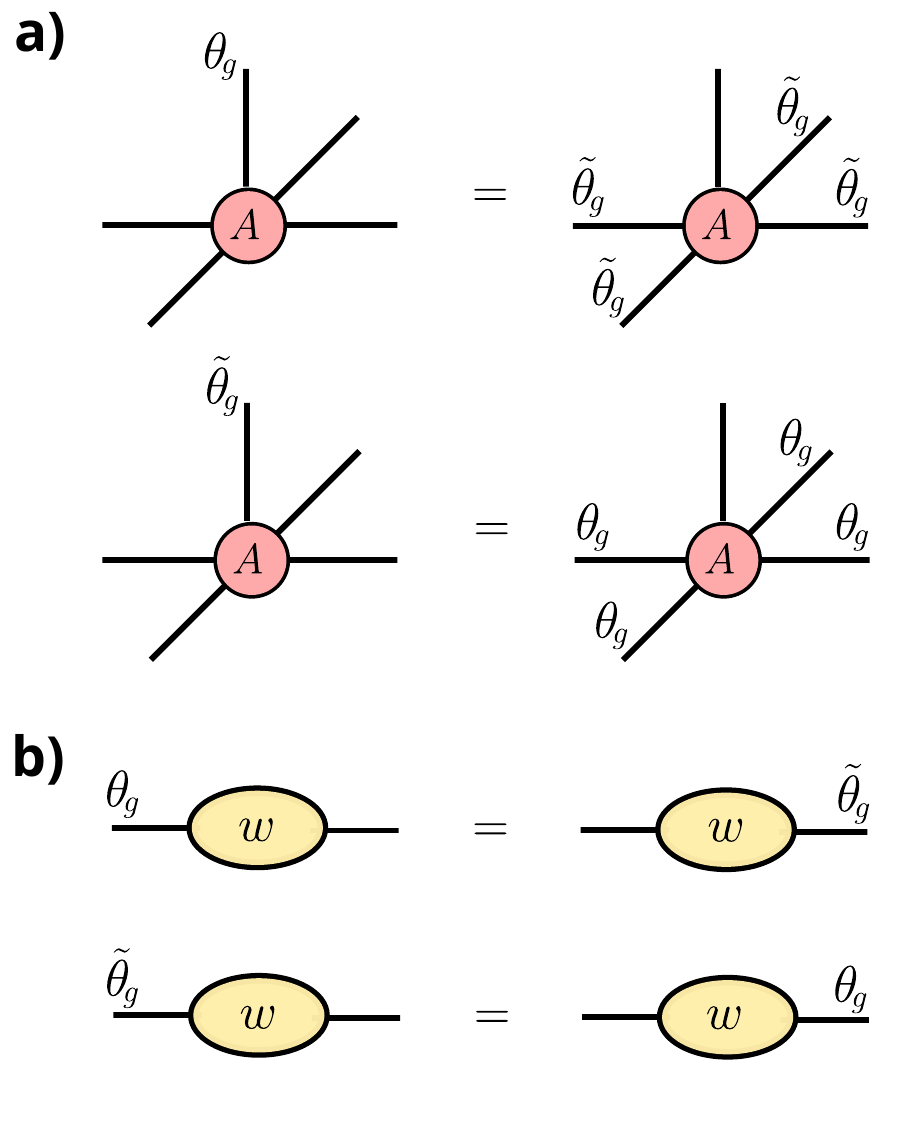}
\captionsetup[subfigure]{labelformat=empty}
    \subfloat[\label{subfig:transformations_a}]{}
    \subfloat[\label{subfig:transformations_b}]{}
    \vspace{-30pt}
	\caption{(a) Graphical representation of the global virtual Gauss law, which relates right and left transformations ($\theta_g$ and $\tilde{\theta}_g$ respectively) on the physical legs of the tensor network to corresponding transformations on the virtual legs. (b) Relation between virtual transformations acting on the left and right legs of the projectors $w$.}
    \label{fig:transformations_ungauged}
\end{figure}

The operators $w\left(\site{x}, k\right)$, defined in equation~\eqref{wdef}, satisfy the invariance properties, or symmetry conditions,
\begin{widetext}
\begin{equation}
\begin{aligned}
\theta^{(J)}_g\left(\site{x},k\right) 
w^{(J)}\left(\site{x},k\right)
\theta_g^{(J)\dagger}\left(\site{x},k\right)=
\widetilde{\theta}_g^{(J)}\left(\site{x}+\hat{\mathbf{e}}_k,k+2\right) 
w^{(J)}\left(\site{x},k\right)
\widetilde{\theta}_g^{(J)\dagger}\left(\site{x}+\hat{\mathbf{e}}_k,k+2\right) ,
\\
\widetilde{\theta}^{(J)}_g\left(\site{x},k\right) 
w^{(J)}\left(\site{x},k\right)
\widetilde{\theta}_g^{(J)\dagger}\left(\site{x},k\right)=
\theta_g^{(J)}\left(\site{x}+\hat{\mathbf{e}}_k,k+2\right) 
w^{(J)}\left(\site{x},k\right)
\theta_g^{(J)\dagger}\left(\site{x}+\hat{\mathbf{e}}_k,k+2\right),
\label{w2trans}
\end{aligned}
\end{equation}
for all sites $\site{x}$, directions $k \in \{1,2\}$, types $J \in \{I, II \}$, and $g \in G$ (as can be seen in Fig.~\ref{subfig:transformations_b}). The link $l = (\site{x}, k)$ is the same as the link $l = (\site{x} + \hat{\mathbf{e}}_k, k+2)$; recall the order of the links shown in figure~\ref{fig:lattice_scheme} (in 3D, this condition is a little more finicky).
For the operators $A(\site{x})$, defined in equation~\eqref{eq:A-def-types}, we must be slightly more careful, because only the type II operators contain physical matter. These obey
\begin{equation}
\begin{aligned}
\theta_g\left(\site{x}\right) 
A^{(II)}\left(\site{x}\right)
\theta^{\dagger}_g\left(\site{x}\right) 
&=
\left[\overset{4}{\underset{k=1}{\prod}}\widetilde{\theta}^{(II)}_g\left(\site{x},k\right)\right]
A^{(II)}\left(\site{x}\right)
\left[\overset{4}{\underset{k=1}{\prod}}\widetilde{\theta}_g^{(II)\dagger}\left(\site{x},k\right)\right],\\
\widetilde{\theta}_g\left(\site{x}\right) 
A^{(II)}\left(\site{x}\right)
\widetilde{\theta}^{\dagger}_g\left(\site{x}\right) 
&=
\left[\overset{4}{\underset{k=1}{\prod}}\theta^{(II)}_g\left(\site{x},k\right)\right]
A^{(II)}\left(\site{x}\right)
\left[\overset{4}{\underset{k=1}{\prod}}\theta_g^{(II)\dagger}\left(\site{x},k\right)\right],
\end{aligned}
\label{globA}
\end{equation}
while the type I operators obey
\begin{equation}
\begin{aligned}
A^{(I)}\left(\site{x}\right)
&=
\left[\overset{4}{\underset{k=1}{\prod}}\widetilde{\theta}^{(I)}_g\left(\site{x},k\right)\right]
A^{(I)}\left(\site{x}\right)
\left[\overset{4}{\underset{k=1}{\prod}}\widetilde{\theta}_g^{(I)\dagger}\left(\site{x},k\right)\right],\\
A^{(I)}\left(\site{x}\right)
&=
\left[\overset{4}{\underset{k=1}{\prod}}\theta^{(I)}_g\left(\site{x},k\right)\right]
A^{(I)}\left(\site{x}\right)
\left[\overset{4}{\underset{k=1}{\prod}}\theta_g^{(I)\dagger}\left(\site{x},k\right)\right],
\end{aligned}
\label{globA1}
\end{equation}
for all sites $\site{x}$ and all $g \in G$ (as can be seen in figure~\ref{subfig:transformations_a}), if the sub-matrix of $\mathcal{T}$ corresponding to type I modes is properly chosen (this equation can also be seen as a requirement for its parameterization).
\end{widetext}

Note that we have reserved $\theta_g(\site{x})$ for the transformations on the physical matter, and $\theta^{(J)}_g(\site{x})$ for the transformations on the virtual modes; we therefore do not define $\theta_g(\site{x}) = \theta^{(I)}_g(\site{x}) \theta^{(II)}_g(\site{x})$ (and similarly for $\tilde{\theta}$). This is purely a matter of notation -- conceptually, it is entirely possible to define an operator as the product of those that act on the type I and type II operators/modes.

\begin{figure}[h]
\includegraphics[width=0.8\linewidth]  {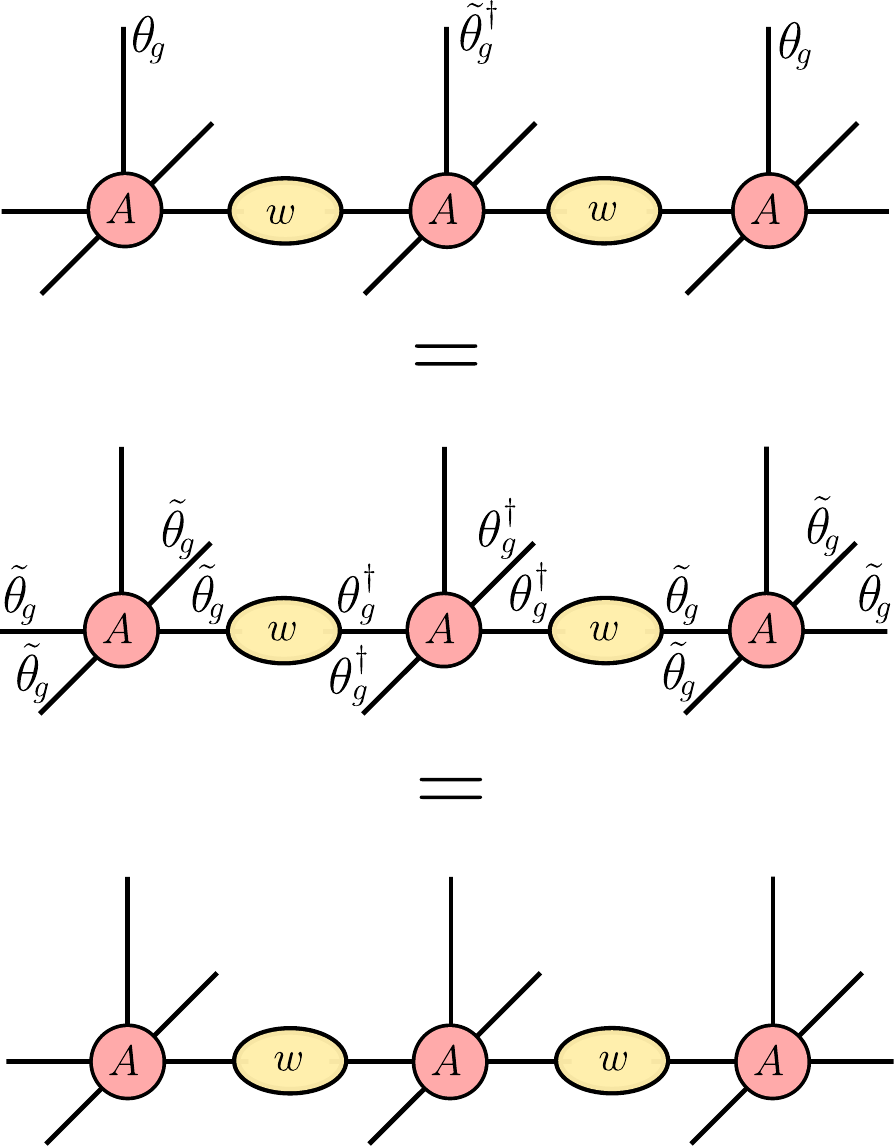}
	\caption{Graphical representation of the global invariance of the PEPS under the action of a transformation belonging to a generic non-Abelian group $G$. The steps follow from the rules shown in figure~\ref{fig:transformations_ungauged}.} The PEPS extends throughout the plane (with periodic boundary conditions), thus the operators on the boundary of the figure also disappear.
    \label{fig:global_invariance}
\end{figure}

As usual with PEPS~\cite{cirac_matrix_2021}, these are the conditions which give rise to a global $G$ symmetry. 
As a result, we deduce that the global invariance of equation~\eqref{zeroinv} holds. 
This is shown graphically in figure~\ref{fig:global_invariance}.

\subsection{Flavors and Chemical Potential}
In cases with multiple fermionic flavors, the above construction generalizes in a straightforward manner.

First, we consider the case when the masses of all flavors are the same.
In the absence of a chemical potential (or in the trivial case in which all flavors are subject to the same chemical potential and the total number of fermions, before the particle-hole transformation, is fixed), the different flavors are not only decoupled, but also contribute to identical Hamiltonians with a flavor permutation symmetry. 
Therefore, if $\ket{\psi_{II}}$ is a ground state of the single flavor Hamiltonian, we  expect copies of the same state, $\bigotimes_f\ket{\psi_{II}}$ to be the ground state of the multi-flavor Hamiltonian.

In the case of different masses and/or different chemical potentials, we must write such a PEPS for each flavor, $\ket{\psi_{II}^{(f)}}$, depending on different parameters, and the state of interest will be $\bigotimes_f\ket{\psi_{II}^{(f)}}$. Such an approach would use new virtual modes for each flavor; one could instead use the same virtual modes.

In particular scenarios there may be additional simplifications.
Consider, for example, the case of two fermionic flavors, $f=1$ and $f=2$, with the same mass but different chemical potentials. 
Since only the difference in the chemical potentials matters, we denote it by $\mu_-$, and write the chemical potential term (before the particle hole transformation) as
\begin{equation}
    H_{\mu} = \frac{\mu_-}{2}\underset{f,\site{x}}{\sum}\left(-1\right)^{f-1}\psi^{\dagger}_{mf}\left(\site{x}\right)\psi_{mf}\left(\site{x}\right).
\end{equation}
Instead of the previous particle-hole transformation, we can perform one which will be carried out on the odd sublattice as before for $f=1$, but on the even sublattice for $f=2$. 
Then, the particle-hole-transformed chemical potential term will take the form
\begin{equation}
    H_{\mu} = \frac{\mu_-}{2}\underset{f,\site{x}}{\sum}\left(-1\right)^{\site{x}}\psi^{\dagger}_{mf}\left(\site{x}\right)\psi_{mf}\left(\site{x}\right).
\end{equation}
The rest of the Hamiltonian is completely invariant under a single site translation, but with this choice, we get a state which is translationally invariant only under translation by two sites. 
Yet the two flavors are still decoupled. 
We can construct the PEPS $\ket{\psi_{II}}$ for $f=1$, which will be a generalization of the case discussed above, for a two-site translation invariance instead of a single one, and then the state of the second flavor will be exactly this state, up to a single site translation. 
Therefore, for all practical purposes it is enough to focus on a single flavor, and simply weaken the translation invariance to be of two sites due to the staggered chemical potential term.

\section{Gauging the PEPS}
PEPS with a global symmetry can easily be gauged following the procedure of~\cite{zohar_building_2016}, which we shall generalize and rephrase here. 
To do so, we use the fact that the invariance conditions of equations~\eqref{globA}  and \eqref{globA1} can be re-written as generalized Gauss' laws, when phrased in terms of the generators:
\begin{equation}
\begin{aligned}
\left[\overset{4}{\underset{k=1}{\sum}}\widetilde{E}_a^{(I)}\left(\site{x},k\right) \right] & A^{(I)}\left(\site{x}\right) \ket{\Omega_{I}} = 0,\\
\left[\overset{4}{\underset{k=1}{\sum}}{E}_a^{(I)}\left(\site{x},k\right) \right] & A^{(I)}\left(\site{x}\right) \ket{\Omega_{I}} = 0,\\
\left[\overset{4}{\underset{k=1}{\sum}}\widetilde{E}_a^{(II)}\left(\site{x},k\right) - Q_a\left(\site{x}\right)\right] & A^{(II)}\left(\site{x}\right) \ket{\Omega_{\text{p}}}\ket{\Omega_{II}} = 0,\\
\left[\overset{4}{\underset{k=1}{\sum}}E_a^{(II)}\left(\site{x},k\right) - \widetilde{Q}_a\left(\site{x}\right)\right] & A^{(II)}\left(\site{x}\right) \ket{\Omega_{\text{p}}}\ket{\Omega_{II}} = 0.
\end{aligned}
\label{virtgauss}
\end{equation}
These resemble the physical Gauss laws from equation~\eqref{Gausslaw}, which gives the \emph{fundamental analogy between PEPS and lattice gauge theories}: the symmetry conditions for a globally invariant PEPS are merely Gauss laws, where the electric fields are virtual. 
In order to obtain a PEPS with a local symmetry, describing the state of a lattice gauge theory, we can begin with a globally invariant PEPS with the same symmetry group, introduce the gauge field Hilbert spaces, and promote the virtual electric fields to physical ones, which lifts the global symmetry to a local one.

To gain some intuition, consider the first line of the virtual Gauss law from equation~\eqref{virtgauss}. 
If we \enquote{replace} the left virtual fields $\widetilde{E}_a$ by the physical left fields $L_a$ on the outgoing links (emanating from the site in the positive directions), and by the physical left fields (due to the minus signs) $-R_a$ on the incoming links (emanating from the site in the negative directions), we will get the physical Gauss law of equation~\eqref{Gausslaw}.

One might worry, however, that this holds only for the even sublattice, since equation~\eqref{Gausslaw} was written before the particle-hole transformation. 
We know that the particle-hole transformation replaces the charge on the odd sublattice with $-\widetilde{Q}_a$; explicitly, the Gauss law, originally specified in equation~\eqref{Gausslaw}, after the particle-hole transformation takes the form
\begin{equation}
\mathsf{D}_a\left(\site{x}\right)\ket{\Psi} = \begin{cases}
                Q_a\left(\site{x}\right)\ket{\Psi} & \site{x} \ \text{even} \\
                -\widetilde{Q}_a\left(\site{x}\right)\ket{\Psi}  & \site{x} \ \text{odd}.
                \end{cases}
\label{newgauss}
\end{equation}
Comparing with the second and fourth lines of equation~\eqref{virtgauss}, we can extend our intuitive guess to \enquote{replace}, on odd sites, the virtual right fields $E_a$ by the physical ones $-L_a$ on the outgoing links, and by $R_a$ on the incoming links.

Before we conclude with a rigorous definition of the gauging, let us complete the discussion by presenting the particle-hole transformed lattice gauge theory Hamiltonian, obtained by performing the transformation of equation~\eqref{parthole} on the full Hamiltonian of equation~\eqref{Hbefore}. 
$H_{\text{E}}$ and $H_{\text{B}}$ are unchanged, as they do not contain the physical fermions. 
$H_{\mu}$ takes the transformed form of the ungauged case, since it only involves same-site terms. 
However, we need to modify the fermionic part, $H_\text{F}$, to
\begin{equation} \begin{aligned}
\label{eq:fermionic-hamiltonian-after-ph}
	H_\text{F} =& \ M\underset{\site{x}}{\sum}\psi_{m}^{\dagger}\left(\site{x}\right)\psi_{m}\left(\site{x}\right) 
 \\ &+ 
		\frac{i}{2a}\bigg(
		 \underset{\site{x}}{\sum} \psi_{m}^{\dagger}\left(\site{x}\right)
   V_{mn}\left(\site{x},1\right)
   \psi^{\dagger}_{n}\left(\site{x}+\hat{\mathbf{e}}_1\right)
	\\ &+
  i \underset{\site{x}}{\sum} \psi_{m}^{\dagger}\left(\site{x}\right)V_{mn}\left(\site{x},2\right)\psi^{\dagger}_{n}\left(\site{x}+\hat{\mathbf{e}}_2\right)
		-\text{h.c.}
		\bigg),
\end{aligned} \end{equation}
where we introduce
\begin{equation} \begin{aligned}
    V_{mn}\left(\site{x},k\right)
    &=
        \begin{cases}
        U_{mn}\left(\site{x},k\right) & \site{x} \ \text{even} \\
        \overline{U}_{mn}\left(\site{x},k\right)   & \site{x} \ \text{odd},
        \end{cases}
        \label{Vdef}
\end{aligned} \end{equation}
for $k\in \{1,2\}$.
The new form of $H_\text{F}$ is invariant under gauge transformation generated by the new Gauss law of equation~\eqref{newgauss}.
One can also see explicitly that a single site translation corresponds to charge conjugation even after the introduction of the gauge field; $V_{mn}\left(\site{x},k\right) \rightarrow V_{mn}\left(\site{x}+\hat{\mathbf{e}}_k,k\right)$ (for $k\in \{1,2\}$) corresponds to $U_{mn}\left(\site{x},k\right) \rightarrow \overline{U}_{mn}\left(\site{x}+\hat{\mathbf{e}}_k,k\right)$, and the pure gauge parts $H_{\text{E}}$ and $H_{\text{B}}$ are trivially invariant. 
Rotation invariance with respect to the rotation defined in equation~\eqref{Urotdef} is also satisfied.

\begin{figure*}[t]
\includegraphics[width=0.8\linewidth]  {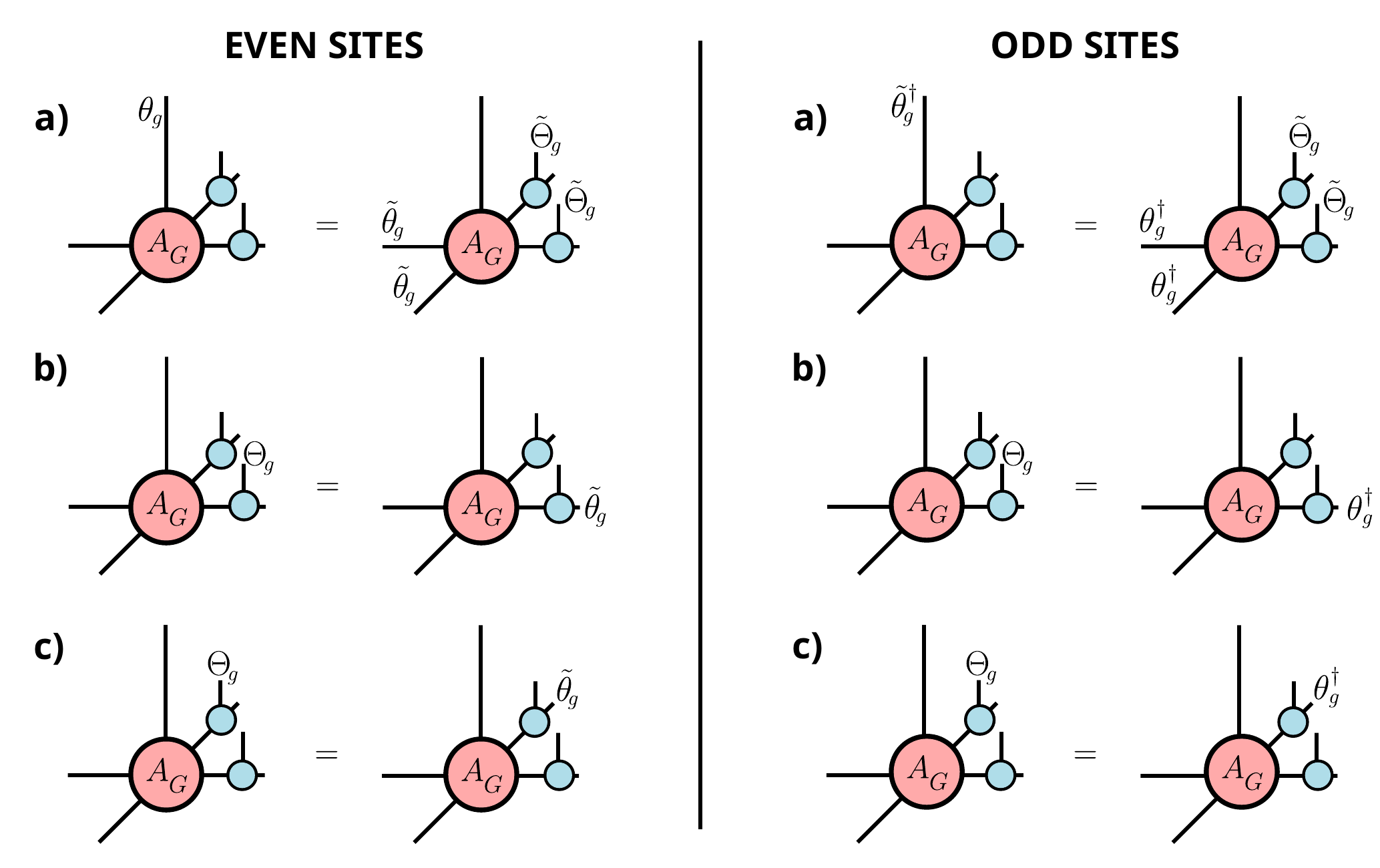}
	\caption{Graphical representation of the relations between physical and virtual local gauge transformation involving both matter and gauge fields. Left and right panels refer to even and odd sites respectively, which differ by the specific form taken by the gauge transformation when acting on the physical matter fields. (a) Once gauge fields are introduced, a physical transformation on the matter fields can be expressed in terms of transformations acting on the virtual \textit{and} gauge fields. (b) and (c) Relations between transformations acting of the gauge fields ($\Theta_g$) and on the  matter fields ($\theta_g$).}
    \label{fig:gauge_transf}
\end{figure*}

\subsection{The Gauging Transformation}
With the intuitive picture in mind, we give a more quantitative and rigorous construction, following the approach of~\cite{zohar_combining_2018,emonts_gauss_2020}. 
First, we introduce on each link a gauge field Hilbert space. 
Note that every link connects two sites, and hence we would like to perform a so-called \emph{gauging transformation} $\mathcal{U}_G$ only on the outgoing modes. 
This transformation will entangle the physical gauge fields with the virtual ones, in a way that will lift the virtual Gauss laws to physical ones.

We introduce the gauging transformation as 
\begin{equation}
\begin{aligned}
\mathcal{U}^{(I)}_G b^{r\dagger}_{k\mu m}\left(\site{x}\right) \mathcal{U}^{(I)\dagger}_G   
&= V^r_{mn}\left(\site{x},k\right)b^{r\dagger}_{k \mu n}\left(\site{x}\right), \\
\mathcal{U}^{(II)}_G c^{\dagger}_{k \mu m}\left(\site{x}\right) \mathcal{U}^{(II)\dagger}_G 
&=
V_{mn}\left(\site{x},k\right)c^{\dagger}_{k \mu n}\left(\site{x}\right), \\ 
\mathcal{U}^{(II)}_G d^{\dagger}_{k \mu m}\left(\site{x}\right) \mathcal{U}^{(II)\dagger}_G 
&=
\overline{V}_{mn}\left(\site{x},k\right)d^{\dagger}_{k \mu n}\left(\site{x}\right),
\end{aligned}
\label{UG2}
\end{equation}
for $k \in \{1,2\}$.
Note that the gauging is done with respect to the gauge group $G$. 
That is, even if the global state $\ket{\psi_0}$ has $U(N)$ invariance, if $G=\SU(N)$, the $U_{mn}$ which we use for gauging will be an element of the gauge group $\SU(N)$, and not $U(N)$. 
Note that this gauging procedure also holds for other gauge groups~\cite{zohar_combining_2018,emonts_gauss_2020}.

The gauging operation acts separately on each link and separately on all the virtual modes on each link, and thus can be factorized as a product of local operations, 
\begin{equation}
    \mathcal{U}^{(J)}_G = \underset{\site{x},k}{\prod}\mathcal{U}^{(J)}_G\left(\site{x},k\right)
\end{equation}
for $k \in \{1,2\}$ and $J\in \{I,II\}$.
Note that if we use the completeness relation of the configuration basis of the gauge field Hilbert space of equation~\eqref{groupcomp}, we can express the gauging operation on a link as
\begin{equation}
    \mathcal{U}^{(J)}_G\left(\site{x},k\right) = 
    \int dg \ket{g}\bra{g}_{\site{x},k} \otimes \mathcal{U}^{(J)}_g \left(\site{x},k\right),
\end{equation}
where
\begin{equation}
    \mathcal{U}^{(J)}_g\left(\site{x},k\right)=     
    \begin{cases}
        \widetilde{\theta}^{(J)}_g\left(\site{x},k\right) & \site{x} \ \text{even} \\
        \theta^{(J)\dagger}_g \left(\site{x},k\right)  & \site{x} \ \text{odd},
    \end{cases}
\end{equation}
for $k\in \{1,2\}$ and $J\in \{I,II\}$. 
We can define
\begin{equation}
    \mathcal{U}_G(\site{x},k) 
    = 
    \mathcal{U}^{(I)}_G(\site{x},k) \ 
    \mathcal{U}^{(II)}_G(\site{x},k).
\end{equation}
$\mathcal{U}_G\left(\site{x},k\right)$ is merely a controlled, entangling operation: the virtual fermions on the outgoing legs are rotated with respect to whatever group element state we have in the physical Hilbert space of the same link.

With this, we define the \emph{gauged} $A^{(J)}$ operators,
\begin{equation}
    A^{(J)}_G\left(\site{x}\right) = 
    \mathcal{U}^{(J)}_G 
    A^{(J)}\left(\site{x}\right)
    \mathcal{U}^{(J)\dagger}_G.
\end{equation}
The global symmetry conditions of equations~\eqref{globA} and~\eqref{globA1} each change into two sets of local symmetry conditions. For the type II operators, on even sites $\site{x}$,
\begin{widetext}
\begin{equation}
\begin{aligned}
\theta_g\left(\site{x}\right) 
A^{(II)}_G\left(\site{x}\right)
\theta^{\dagger}_g\left(\site{x}\right) 
&=
\left[\overset{2}{\underset{i=1}{\prod}}\widetilde{\theta}^{(II)}_g\left(\site{x},i+2\right)\widetilde{\Theta}_g\left(\site{x},i\right)\right]
A^{(II)}_G\left(\site{x}\right)
\left[\overset{2}{\underset{i=1}{\prod}}\widetilde{\theta}^{(II)\dagger}_g\left(\site{x},i+2\right)\widetilde{\Theta}^{\dagger}_g\left(\site{x},i\right)\right],\\
\Theta_g\left(\site{x},k\right) 
A^{(II)}_G\left(\site{x}\right)
\Theta^{\dagger}_g\left(\site{x},k\right) 
&=
\widetilde{\theta}^{(II)}_g\left(\site{x},k\right)
A^{(II)}_G\left(\site{x}\right)
\widetilde{\theta}^{(II)\dagger}_g\left(\site{x},k\right),
\end{aligned}
\label{locAeven}
\end{equation}
for $k\in \{1,2\}$ and any $g \in G$ (as can be seen in figure~\ref{fig:gauge_transf} in the left panel). 
On odd sites $\site{x}$,\begin{equation}
\begin{aligned}
\widetilde{\theta}^{\dagger}_g\left(\site{x}\right) 
A^{(II)}_G\left(\site{x}\right)
\widetilde{\theta}_g\left(\site{x}\right) 
&=
\left[\overset{2}{\underset{i=1}{\prod}}\theta^{(II)\dagger}_g\left(\site{x},i+2\right)\widetilde{\Theta}_g\left(\site{x},i\right)\right]
A^{(II)}_G\left(\site{x}\right)
\left[\overset{2}{\underset{i=1}{\prod}}\theta^{(II)}_g\left(\site{x},i+2\right)\widetilde{\Theta}^{\dagger}_g\left(\site{x},i\right)\right],\\
\Theta_g\left(\site{x},k\right) 
A^{(II)}_G\left(\site{x}\right)
\Theta^{\dagger}_g\left(\site{x},k\right) 
&=
\theta^{(II)\dagger}_g\left(\site{x},k\right)
A^{(II)}_G\left(\site{x}\right)
\theta^{(II)}_g\left(\site{x},k\right),
\end{aligned}
\label{locAodd}
\end{equation}
for $k\in \{1,2\}$ and any $g \in G$ (as can be seen in figure~\ref{fig:gauge_transf} in the right panel).

For the type I operators, we have the same relations, but without the operators on the physical matter,
\begin{equation}
\begin{aligned}
A^{(I)}_G\left(\site{x}\right)
&=
\left[\overset{2}{\underset{i=1}{\prod}}\widetilde{\theta}^{(I)}_g\left(\site{x},i+2\right)\widetilde{\Theta}_g\left(\site{x},i\right)\right]
A^{(I)}_G\left(\site{x}\right)
\left[\overset{2}{\underset{i=1}{\prod}}\widetilde{\theta}^{(I)\dagger}_g\left(\site{x},i+2\right)\widetilde{\Theta}^{\dagger}_g\left(\site{x},i\right)\right],\\
\Theta_g\left(\site{x},k\right) 
A^{(I)}_G\left(\site{x}\right)
\Theta^{\dagger}_g\left(\site{x},k\right) 
&=
\widetilde{\theta}^{(I)}_g\left(\site{x},k\right)
A^{(I)}_G\left(\site{x}\right)
\widetilde{\theta}^{(I)\dagger}_g\left(\site{x},k\right),
\end{aligned}
\end{equation}
for $k\in \{1,2\}$ and any $g \in G$ (as can be seen in figure~\ref{fig:gauge_transf} in the left panel). 
On odd sites $\site{x}$,\begin{equation}
\begin{aligned}
A^{(I)}_G\left(\site{x}\right)
&=
\left[\overset{2}{\underset{i=1}{\prod}}\theta^{(I)\dagger}_g\left(\site{x},i+2\right)\widetilde{\Theta}_g\left(\site{x},i\right)\right]
A^{(I)}_G\left(\site{x}\right)
\left[\overset{2}{\underset{i=1}{\prod}}\theta^{(I)}_g\left(\site{x},i+2\right)\widetilde{\Theta}^{\dagger}_g\left(\site{x},i\right)\right],\\
\Theta_g\left(\site{x},k\right) 
A^{(I)}_G\left(\site{x}\right)
\Theta^{\dagger}_g\left(\site{x},k\right) 
&=
\theta^{(I)\dagger}_g\left(\site{x},k\right)
A^{(I)}_G\left(\site{x}\right)
\theta^{(I)}_g\left(\site{x},k\right),
\end{aligned}
\end{equation}
for $k\in\{1,2\}$ and any $g \in G$ (as can be seen in figure~\ref{fig:gauge_transf} in the right panel).
\end{widetext}

\begin{figure*}[t]
\includegraphics[width=0.9\linewidth]  {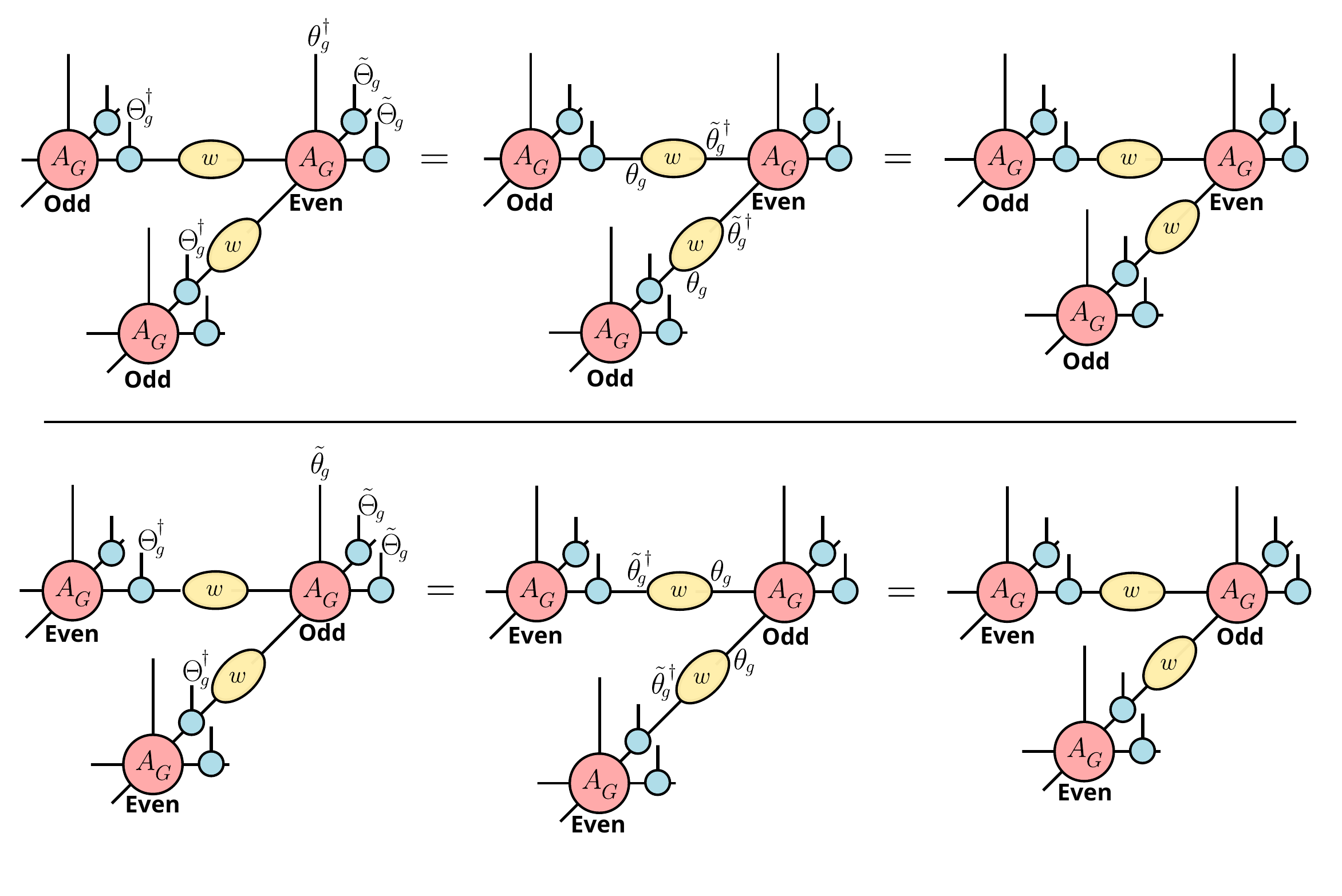}
	\caption{Graphical demonstration of the local invariance of
the PEPS under a gauge transformation on even (upper panel) and odd (lower panel) sites. The steps follow from the rules shown in figures~\ref{fig:transformations_ungauged} and~\ref{fig:gauge_transf}.}
\label{fig:gauge_invariance}
\end{figure*}

The gauging operators $\mathcal{U}_G$ act on the gauge field Hilbert space.
Therefore, to finish the construction, we must consider the state of the gauge fields. 
Let $\ket{\text{in}}$ be some gauge field state which is invariant under pure gauge transformations, that is\begin{equation}
\mathsf{D}_a\left(\site{x}\right)\ket{\text{in}} = 0
\label{Dzero}
\end{equation}
for all the generators of the gauge group, and any site $\site{x}$ (the divergence of the electric fields, $\mathsf{D}_a\left(\site{x}\right)$, was defined in equation~\eqref{Gausslaw} and is given after the particle hole transformation by~\eqref{newgauss}). 
Note that we can encode states in static charge sectors other than that of equation~\eqref{Dzero} by properly choosing $\ket{\text{in}} $ to be in the appropriate sector. 
Below we will focus on the cases where equation~\eqref{Dzero} is satisfied.

Since $\mathcal{U}_G\ket{\Omega_{II}}=\ket{\Omega_{II}}$, the state
\begin{equation}
\ket{\Psi}=\bra{\Omega_{II}} \underset{\site{x},k}{\prod}w^{(II)\dagger}\left(\site{x},k\right)
\mathcal{U}_G^{(II)}
\underset{\site{x}}{\prod}A^{(II)}\left(\site{x}\right)\ket{\Omega_{\text{p}}}\ket{\Omega_{II}} \ket{\text{in}}
\label{GPEPSdef}
\end{equation}
is gauge invariant under $G$. 
That is, it satisfies the particle-hole transformed Gauss law of equation~\eqref{newgauss}.

Since the finite transformation version holds too, the above construction will hold for finite groups as well. For this reason, the transformation properties of the PEPS ingredients were given in a finite transformation form and not in terms of generators. The generalization of generator equations, such as equation \eqref{Dzero}, is straightforward.

A graphical demonstration of the invariance under gauge transformations is given in figure~\ref{fig:gauge_invariance}.

The state $\ket{\Psi}$ is a \emph{gauged Gaussian fermionic PEPS} if the initial gauge field state $\ket{\text{in}}$ is also a PEPS. In section~\ref{sec:group-element-expansion}, we discuss various options for the state $\ket{\text{in}}$, and show how it relates to the amplitudes $\psi_I$.

\vspace{1em}

\subsection{The Group Element Basis Expansion}
\label{sec:group-element-expansion}
Given a gauge field configuration state $\ket{\mathcal{G}}$, as defined in equation~\eqref{confdef}, we can also introduce the notation $\mathcal{U}^{(J)}_G\left[\mathcal{G}\right]$, as the product of $\mathcal{U}^{(J)}_g\left(\site{x},k\right)$ operators for all the group elements contained in $\ket{\mathcal{G}}$. 
This allows us, using the completeness relation of the configuration basis of the whole lattice from equation (\ref{compconf}), to express the gauged Gaussian fermionic PEPS in the configuration basis expansion of equation~\eqref{Gexpansion}, where
\begin{widetext}
\begin{equation}
\ket{\psi_{II}\left(\mathcal{G}\right)}=\bra{\Omega_{II}} \underset{\site{x},k}{\prod}w^{(II)\dagger}\left(\site{x}, k\right)
\mathcal{U}^{(II)}_G \left[\mathcal{G}\right]
\underset{\site{x}}{\prod}A^{(II)}\left(\site{x}\right)\ket{\Omega_{\text{p}}}\ket{\Omega_{II}} 
\end{equation}
is indeed a Gaussian fermionic PEPS, allowing for efficient computations as desired.

If we have several flavors, $f\in \{1,...,F\}$, we can take the state to be
\begin{equation}
\ket{\psi_{II}\left(\mathcal{G}\right)}= \overset{F}{\underset{f=1}{\bigotimes}}\ket{\psi^{(f)}_{II}\left(\mathcal{G}\right)},
\label{psi2F}
\end{equation}
where for each flavor we construct a Gaussian PEPS separately,
\begin{equation}
\label{eq:flavor-peps}
\ket{\psi^{(f)}_{II}\left(\mathcal{G}\right)}=    
\bra{\Omega^{(f)}_{II}} \underset{\site{x},k}{\prod}w^{(f,II)\dagger}_k\left(\site{x}\right)
\mathcal{U}^{(f,II)}_G \left[\mathcal{G}\right]
\underset{\site{x}}{\prod}A^{(f,II)}\left(\site{x}\right)\ket{\Omega^{(f)}_{\text{p}}}\ket{\Omega^{(f)}_{II}} 
\end{equation}
\end{widetext}
where all the ingredients but the gauge field are defined for each flavor separately. 
This is possible since there is no direct coupling between the different flavors, even after gauging; the only coupling is mediated through the gauge field, where all flavors experience the same $\mathcal{G}$ configuration (gauging is flavor-independent). Therefore, for a given gauge field configuration $\mathcal{G}$, we can expand the matter as a product of independent states.
As noted above, it is also possible to ``reuse" the type II virtual modes for additional flavors; if one wishes, one can enforce a permutation symmetry by imposing such a symmetry on $\mathcal{T}$ and $W$. 

Comparing equations~\eqref{Gexpansion} and~\eqref{GPEPSdef}, we see that $\ket{\text{in}} = \int \mathcal{DG} \ \psi_I(\mathcal{G}) \ket{\mathcal{G}}$, or equivalently,
\begin{equation}
    \psi_{I}\left(\mathcal{G}\right) =
    \braket{\mathcal{G}}{\text{in}}.
\end{equation}
There are several options available for the $\ket{\text{in}} $ states. 

Before showing how the choice of $\ket{\text{in}}$ fits in to the PEPS construction above, we can first consider the trivial option,
\begin{equation}
    \ket{\text{in}} = \underset{\site{x},k=1,2}{\bigotimes}\ket{000}
    \equiv \ket{\mathbf{0}}
\end{equation}
where $\ket{000}$ is the singlet ($j=0$) state, invariant under all group transformations $\Theta_g$ and $\widetilde{\Theta}_g$ (i.e. the zero electric field state).

In this case, using the basis change  formula of equation~\eqref{peterweyl}, we get that
$\psi_{I}\left(\mathcal{G}\right)$ is a constant, and since the state is not normalized in any case, we can set it to 1. 
This is the option chosen in references~\cite{zohar_fermionic_2015,zohar_projected_2016,zohar_combining_2018}, and it involves no computational challenge at all.

Alternatively, since the only requirement for the $\ket{\Psi}$ of equation~\eqref{GPEPSdef} to be a gauge invariant state is that  $\ket{\text{in}}$ be pure-gauge invariant, we can construct such a state in any way we wish. 
In our case, it makes sense to construct it as a PEPS, as this gives the desired entanglement properties while maintaining the easy symmetry encoding. 
This guarantees that the whole state $\ket{\Psi}$ is  a PEPS.
If we further make it a gauged Gaussian PEPS, we can use a lot of the machinery introduced above, as already explained, and can perform calculations with it efficiently using the Gaussian formalism.
This explains the introduction of the type I modes in section~\ref{sec:peps-construction-main} - they can be used to define the state $\ket{\text{in}}$.
That is, 
\begin{equation}
\psi_{I}\left(\mathcal{G}\right) =   \bra{\Omega_{I}} \underset{\site{x},k}{\prod}w^{(I)\dagger}\left(\site{x}, k\right) \mathcal{U}^{(I)}_G \left[\mathcal{G}\right]  \underset{\site{x}}{\prod}A^{(I)}\left(\site{x}\right)  \ket{\Omega^{(l)}_{I}},
\label{psiIdef}
\end{equation}
and plugging this in to equation~\eqref{GPEPSdef}, we have
\begin{equation} \label{eq:full-state-encapsulated}
    \ket{\Psi}  = \int \mathcal{DG} \ 
    \psi_I(\mathcal{G}) \ket{\mathcal{G}} \ket{\psi_{II}(\mathcal{G})},
\end{equation}
where we have chosen $\ket{\mathbf{0}}$ as the state on which to build $\psi_I(\mathcal{G})$.

\subsection{The Full Ansatz}

We can write the final state in a variety of equivalent ways. We have already done so at various levels of detail, as can be seen from equations~~\eqref{GPEPSdef} and~\eqref{eq:full-state-encapsulated}.
Here, we present one final unified way of writing the state, and then expand it in a slightly new way that is particularly convenient for improving the efficiency of computations, which we discuss more in the next section.

Rewriting equation~\eqref{eq:full-state-encapsulated}, we have
\begin{equation} \label{eq:full-ansatz-final}
    \ket{\Psi} = \bra{\Omega_{\text{v}}}
   \underset{\site{x},k}{\prod}w^{\dagger}\left(\site{x},k\right)
   \mathcal{U}_G 
\underset{\site{x}}{\prod}A\left(\site{x}\right)  \ket{\Omega_{\text{v}}}\ket{\Omega_{\text{p}}}\ket{\mathbf{0}}
\end{equation}
where the Fock vacua and all the operators include all physical and virtual modes, and $\mathcal{U}_G = \mathcal{U}^{(I)}_G \mathcal{U}^{(II)}_G$.

As has been discussed, we can divide $\ket{\Psi}$ into two, corresponding to a state of the matter $\ket{\psi_{II}\left(\mathcal{G}\right)}$ and an amplitude $\psi_I(\mathcal{G})$.
Further, when considering multiple fermionic flavors, $\ket{\psi_{II}(\mathcal{G})}$ can be written as a product of PEPS as was shown in equation~\eqref{psi2F}. 
These are both examples of a more general principle. So long as 
\begin{enumerate}
\item the full $\mathcal{T}$ matrix as seen in equation~\eqref{eq:T-mat-illustration}, which enters into the full $A(\site{x})$ of equation~\eqref{eq:A-op-combined}, is block diagonal (this may be easier to see if the modes are reordered);
\item the projectors $w(\site{x}, k)$ do not mix the different blocks;
\end{enumerate}
we can write the full ansatz as a product of separate PEPS. 
We refer to each PEPS as a \textit{layer}; of course, each layer can only contain virtual modes of either type I or type II, but not both. The division into layers forms a partition of all the modes $\{\Psi_\alpha^\dagger \}$.

$\ket{\psi_{II}(\mathcal{G})}$ has already been given this form -- written as a product of layers, each involving modes that do not interact -- in equation~\eqref{psi2F}. 

We can also introduce separate \emph{layers} of type I modes -- that is, sets of copies which are uncoupled by the $A^{(I)}$ operators, similarly to the flavors in the type II case. 
In the type II case, this had a physical meaning, but in the type I case it may be seen as a numerical trick; though it may be used, for example, to separate between the type I fermions and the bosons which cannot be directly coupled. 
If we have several layers labeled by $l \in \{1,...,L\}$, we can express
\begin{equation}
   \psi_{I}\left(\mathcal{G}\right) =\overset{L}{\underset{l=1}{\prod}}\psi^{(l)}_{I}\left(\mathcal{G}\right).
\end{equation}
where
\begin{equation}
\psi^{(l)}_{I}\left(\mathcal{G}\right) =   \bra{\Omega^{(l)}_{I}} \underset{\site{x},k}{\prod}w^{(l,I)\dagger}\left(\site{x}, k\right) \mathcal{U}^{(l,I)}_G \left[\mathcal{G}\right]  \underset{\site{x}}{\prod}A^{(l,I)}\left(\site{x}\right)  \ket{\Omega^{(l)}_{I}}
\label{psiIdef-per-layer}
\end{equation}
and the only relation between the different layers is that they are coupled to the same gauge field. 
This was successfully used for the study of the pure gauge $\mathbb{Z}_2$ LGT in~\cite{emonts_finding_2023}.

When pure gauge theories are considered, we do not need the type II modes at all. 
On the other hand, we could build an equivalent state, as done in previous works~\cite{zohar_fermionic_2015,zohar_projected_2016}  by setting the coupling of type II fermions with physical fermions to zero, and choosing the trivial state $\ket{\text{in}}$. 
In such a case, the distinction between types I and II almost disappears. It nevertheless remains important, 
primarily because type II fermions carry many more charges due to their coupling to the physical fermions, which imposes some spin, doubling, and color restrictions. 

To finish this section, note that each copy of type II modes ($c^\dagger$ or $d^\dagger$) has $\dim(j)$ colors (values over which $m$ ranges). Each copy of the $b^\dagger$ modes may, as discussed, have a different representation $r$, and so has $\dim(r)$ colors. If we denote the number of type I copies by $N_b$ and the number of type II copies by $N_{c,d}$, then the total number of virtual modes is $2d [N_b \dim(r) + N_{c,d}\dim(j)]$ per site, as each site is connected to $2d$ links.

\section{The Algorithm}
\label{sec:algorithm}

As explained in section~\ref{sectionwhy}, we are interested in computing observables with respect to our gauged Gaussian PEPS $\ket{\Psi}$ using Monte-Carlo sampling of the gauge fields in the configuration (group element) basis. 
Furthermore, we can use this as a building block for a variational Monte-Carlo ground state search, by minimizing the expectation value of the Hamiltonian with respect to the parameters on which $\ket{\Psi}$ depends (while there are many choices available in constructing the ansatz, the only variational parameters are those in $\mathcal{T}$).
In this section we elaborate on the computation of the expectation value of an observable, given some fixed choice of the parameters, using the Monte-Carlo method.

To do this, we first expand our state in the configuration basis. We assume that we have $F$ physical flavors and $L$ layers of type I modes. 
Then, the ansatz state will take the form
\begin{equation}
    \ket{\Psi} = \int \mathcal{DG} \ket{\mathcal{G}}
     \overset{L}{\underset{l=1}{\prod}}\psi^{(l)}_{I}\left(\mathcal{G}\right)
     \overset{F}{\underset{f=1}{\bigotimes}}\ket{\psi^{(f)}_{II}\left(\mathcal{G}\right)}.
     \label{fullPsi}
\end{equation}

In the next two subsections, we discuss how this state can be used in numerical calculations.

\subsection{Evaluation of the Integrands}
The expectation value of any gauge invariant observable can be written
\begin{equation} \label{eq:obs-expectation}
\left\langle \mathcal{O} \right\rangle = \int \mathcal{DG} \ F_{\mathcal{O}}\left(\mathcal{G}\right)p\left(\mathcal{G}\right).
\end{equation}
We begin by discussing the evaluation of the integrands; the evaluation of the integral over gauge field configurations is discussed in the next subsection. 

The function $F_{\mathcal{O}}(\mathcal{G})$ can be calculated from the covariance matrices of the state and its components.
In practice, the division of the state into layers, whether of type I or type II, which are all of the form of equations~\eqref{eq:flavor-peps} or \eqref{psiIdef-per-layer}, allows the calculation of $F_{\mathcal{O}}(\mathcal{G})$ through separate calculations for each layer. The results can then be combined to give the expectation value of the full state. This requires one to have an explicit expression for $F_{\mathcal{O}}(\mathcal{G})$ in terms of the covariance matrix elements, which will determine how to combine the results from the various layers. 

Separating the calculation across different layers provides a significant computational benefit. Since we use covariance matrices in all calculations, the size of those matrices is, in practice, a major factor in the time required for calculations (though the scaling remains polynomial). The separation into layers means that the covariance matrices can be calculated for each layer separately, and never need to be combined -- the size of the covariance matrices is determined for each layer by the number of modes (physical and virtual) in that layer.

For instance, consider the mesonic operator defined in equation~\eqref{mesdef}, which can be calculated using elements of the covariance matrix of the type II fermions. 
Note that both equations need to undergo the appropriate particle-hole transformation of equation~\eqref{parthole}, depending on the sublattices of the meson's beginning, $\site{x}$, and end, $\site{y}$. 
For example, if $\site{x}$ is on the even sublattice and $\site{y}$ on the odd one, the mesonic operator of equation~\eqref{mesdef} takes the form
\begin{equation}
    \mathcal{M}_f\left(\site{x},\mathcal{C},\site{y}\right) = 
    \psi^{\dagger}_{mf}\left(\site{x}\right)
    \left[\mathcal{P}\underset{\ell \in \mathcal{C}}{\prod}U\left(\ell\right)\right]_{mn}
    \psi^{\dagger}_{nf}\left(\site{y}\right),   
\end{equation}
and its expectation value is given by
\begin{widetext}
\begin{equation} \begin{aligned}
    \langle & \mathcal{M}_f \left(\site{x},\mathcal{C},\site{y}\right) \rangle 
    & =
    \int \mathcal{DG} \ \left[\mathcal{P}\underset{\ell \in \mathcal{C}}{\prod}D\left(\ell\right)\right]_{mn}
    \frac{
    \bra{\psi\left(\mathcal{G}\right)} \psi^{\dagger}_{mf}\left(\site{x}\right)
        \psi^{\dagger}_{nf}\left(\site{y}\right) \ket{\psi(\mathcal{G})}}
        {\braket{\psi (\mathcal{G})}}
                p\left(\mathcal{G}\right).
        \label{mestrans}
\end{aligned} \end{equation}
Taking advantage of the fact that the fermionic operators do not act on $\psi_I(\mathcal{G})$, it can be seen from the flavor decomposition of equation~\eqref{fullPsi} that
\begin{equation}
    \frac{
    \bra{\psi\left(\mathcal{G}\right)} \psi^{\dagger}_{mf}\left(\site{x}\right)
        \psi^{\dagger}_{nf}\left(\site{y}\right) \ket{\psi\left(\mathcal{G}\right)}}
        {\braket{\psi\left(\mathcal{G}\right)}}
    =
    \frac{
        \bra{\psi^{(f)}_{II}\left(\mathcal{G}\right)} \psi^{\dagger}_{mf}\left(\site{x}\right)
        \psi^{\dagger}_{nf}\left(\site{y}\right) \ket{\psi^{(f)}_{II}\left(\mathcal{G}\right)}}
        {\braket{\psi^{(f)}_{II}\left(\mathcal{G}\right)}},
\end{equation}
\end{widetext}
which will depend only on the covariance matrix of $\ket{\psi^{(f)}_{II}\left(\mathcal{G}\right)}$. Other operators will in general have contributions from multiple layers (for example, gradients depend on the norm of the state, which  depends on all layers).
        
For operators depending on the electric fields, which are not diagonal in the group element basis, such as those contained in $H_{\text{E}}$, we need to use the formula of equation~\eqref{eintdef}, which does not change under the particle hole transformation. 
These expectation values depend on overlaps of Gaussian states and wave functions which can also be factorized by layer (including flavor), using the decomposition of equation~\eqref{fullPsi}. 
For each layer, it is possible to use a trick introduced and demonstrated in~\cite{emonts_finding_2023}, in which some virtual modes (whether type I or type II), on the particular link where the evaluation takes place, are treated as ``physical" and the required information can be extracted efficiently from their covariance matrix (this allows one to avoid the numerical bottlenecks imposed by Pfaffians encountered in~\cite{emonts_variational_2020}). 
An example for the $\mathbb{Z}_2$ electric operator can be seen in~\cite{emonts_finding_2023}. 

As explained in section~\ref{sectionwhy}, and as can be seen in equation~\eqref{eq:obs-expectation}, the probability density $p\left(\mathcal{G}\right)$ defined in equation~\eqref{pdef} is a crucial component of any Monte Carlo calculation.
The better quantity to compute for the Monte-Carlo sampling, however, is the unnormalized probability, $p_0\left(\mathcal{G}\right)$, defined in the same equation; as arises from equation~\eqref{eq:acceptance-probability} below, the full probability is not necessary. 
Using the expansion of equation~\eqref{fullPsi}, we can write it as
\begin{equation}
p_0\left(\mathcal{G}\right) = 
\overset{L}{\underset{l=1}{\prod}}  \left|\psi^{(l)}_{I}\left(\mathcal{G}\right)\right|^2 
\overset{F}{\underset{f=1}{\prod}} \braket{\psi^{(f)}_{II}\left(\mathcal{G}\right)}.
\end{equation}
For any number of layers and flavors, the unnormalized probability is a product of squared norms of Gaussian states, which can be computed separately and independently, reducing the size of matrices we have to deal with.

We leave the explicit form of $F_{\mathcal{O}}(\mathcal{G})$ for any given operator $\mathcal{O}$ to other work. 
In appendix~\ref{sec:covariance} we elaborate on the calculation of the covariance matrices, and show how the norm -- which gives $p(\mathcal{G})$ -- can be calculated.

\subsection{Evaluation of the Integrals}

\subsubsection{The Monte-Carlo Procedure}

To calculate the full expectation value of an observable, we must integrate the integrands over all the gauge field configurations. 
In practice, for all but the simplest systems, this requires Monte Carlo evaluation. 
We follow the prescription for Markov Chain Monte Carlo (MCMC)~\cite{metropolis_equation_1953,creutz_monte_1980,creutz_monte_1983}.
This section provides no new information to one who is familiar with such methods.

In accordance with the MCMC proecure we  start with a warmup phase: pick a random gauge configuration $\mathcal{G}$, build the covariance matrices, and evaluate the probability $p_0(\mathcal{G})$. Repeatedly pick a new $\mathcal{G}'$ by choosing a random fixed number of links and updating the gauge fields to a randomly chosen value on each, and accepting or rejecting the update with probability
\begin{equation} \label{eq:acceptance-probability}
\min\left(\frac{p(\mathcal{G}')}{p(\mathcal{G})}, 1\right) = \min\left(\frac{p_0(\mathcal{G}')}{p_0(\mathcal{G})}, 1\right). 
\end{equation} 
When convergence to $p(\mathcal{G})$ is reached, move to the next step, the measurement of observables.
In the measurement phase,  continually update the gauge field configuration as above, but each time an update is accepted or rejected, compute the value of any observables (including the combinations from all layers), and store the result together with the probability.
Note that if the proposed $\mathcal{G}'$ is not accepted, we resample $\mathcal{G}$, and include its measurement contribution an additional time.

In both of these phases, for each accepted configuration $\mathcal{G}$, we calculate the value of the observable (or just the probability during warmup) for the state given that configuration.
Thus, for each new gauge field configuration, we need to reconstruct those covariance matrices which depend on the gauge field configuration (see appendix~\ref{sec:covariance} for details). 
Upon sampling a new gauge field configuration, it is not necessary to recalculate the covariance matrices from scratch; local updates can be applied based on the links whose gauge values were changed, which provides significant efficiency gains. 
More details are given in appendix~\ref{sec:local_updates}.

Finally, calculate expectation values: after sufficient many samples are collected (as determined, e.g., by the expected error), compute the expectation value by taking an average of the calculated values weighted by the probabilities.

One application of this procedure is finding the ground state of a system of interest through variational Monte Carlo, as done in~\cite{emonts_variational_2020,emonts_finding_2023}.
To do so, one must start with a guess for all the state parameters (a random guess is often the best one can do), and evaluate the gradients of the Hamiltonian using the Monte Carlo procedure just described. 
After the evaluation of the gradients, the parameters can be updated in the direction that minimizes the energy; the full variational search continues with repeated gradient evaluations and updates until some convergence condition is reached.

The state as a whole, as well as each layer, is composed out of three types of operators: $w, \mathcal{U}_G$, and $A$. 
While there are some choices (discussed above) that go into the construction of $w$, these determine the form of the ansatz, and $w$ does not contain variational parameters. 
$\mathcal{U}_G$ is determined by the gauge group and the number and types of virtual modes, and also does not contain any variational freedom. 
Thus, all the variational parameters are left in $A$ -- in particular, in $\mathcal{T}$. 
To find the ground state, one therefore performs gradient descent, or a similar algorithm, on these parameters, where at each step one uses the Monte Carlo procedure to evaluate gradients.

\subsubsection{Gauge Fixing}
\label{sec:gauge-fixing}
Since the ansatz state $\ket{\Psi}$ is gauge invariant, we can use gauge fixing in order to reduce the number of gauge field configuration samples required.

All  observables $\mathcal{O}$ of interest are gauge invariant. 
We compute their expectation values using integrals over all the gauge field configurations, through equation~\eqref{eq:obs-expectation}.
The integrands of all these integrals are \emph{pure-gauge invariant}, as we now discuss.

We begin with the probability density, $p\left(\mathcal{G}\right)$. 
To prove its gauge invariance, it is enough to prove the gauge invariance of the unnormalized version. 
It is defined as the norm of some state of matter experiencing a gauge field configuration $\mathcal{G}$, 
\begin{equation}
    p_0\left(\mathcal{G}\right) = \braket{\psi\left(\mathcal{G}\right)}.
\end{equation}
If $\mathcal{G}'$ is equivalent, through some pure-gauge transformation, to $\mathcal{G}$, there exists some unitary transformation $\hat{\mathcal{U}}$ acting locally on the matter, such that
\begin{equation}
    \hat{\mathcal{U}}\ket{\psi\left(\mathcal{G}\right)} = \ket{\psi\left(\mathcal{G}'\right)},
\end{equation}
(this is the unitary which completes the pure-gauge transformation to a full gauge transformation).
Since unitary transformations preserve the norms, we deduce that 
\begin{equation}
p_0\left(\mathcal{G}'\right)=\braket{\psi\left(\mathcal{G}'\right)}  
=\bra{\psi\left(\mathcal{G}\right)}\hat{\mathcal{U}}^{\dagger}\hat{\mathcal{U}}\ket{\psi\left(\mathcal{G}\right)}=p_0\left(\mathcal{G}\right).
\end{equation}

If the gauge field operator $\mathcal{O}$ depends on the gauge field alone, such as in the cases of a flux loop or a function of the electric fields, the corresponding function $F_{\mathcal{O}}\left(\mathcal{G}\right)$ will be pure-gauge invariant, since it is a gauge invariant function of the gauge field configurations alone.

If we are interested in computing the expectation values of a mesonic operator, after the particle-hole transformation, we must compute an integral like that of equation~\eqref{mestrans}. 
The corresponding $F_{\mathcal{O}}\left(\mathcal{G}\right)$ are gauge invariant too; the change in $\left[{\prod}D\left(\ell\right)\right]_{mn}$ which occurs when transforming $\mathcal{G}$ to $\mathcal{G}'$ is compensated by the matter transformations $\hat{U}$ which transforms the fermionic operators; in other words, the transformations of the string and the covariance matrix compensate for each other. 

Therefore, unsurprisingly, for all gauge invariant operators, 
\begin{equation}
F_{\mathcal{O}}\left(   \mathcal{G}'\right) =  F_{\mathcal{O}}\left(   \mathcal{G}\right),
\end{equation}
and the integrands are gauge invariant.
Hence, for any $\mathcal{G}'$  which is equivalent, through some gauge transformation, to $\mathcal{G}$, we have\begin{equation}
\left\langle \mathcal{O} \right\rangle = \int \mathcal{DG} \ F_{\mathcal{O}}\left(\mathcal{G}'\right)p\left(\mathcal{G}'\right)
\end{equation}
which does not require changing the integration variable to $\mathcal{G}'$.

Let $\mathbf{T}$ be some maximal tree (an open path without closed loops with a maximal number of links) on our lattice. 
For any $\mathcal{G}$, we can find a gauge equivalent $\mathcal{G'}$ for which all the links on the maximal tree will be  fixed to the identity of the group, which makes the integration over these links redundant. 
We deduce that one does not need to integrate over these links; and, since we set them to the group identity, they do not have any effect through gauging (i.e. we can simply not gauge the links on $\mathbf{T}$ when computing the norms and the covariance matrices).

This will significantly reduce the number of gauge field configurations over which one needs to integrate. Consider, for example, $d=2$ with
open boundary conditions: a maximal tree could be, for example, all the horizontal links, and the vertical ones along one column. 
For periodic boundary conditions, a maximal tree could include all the horizontal links but one on each row, and all the vertical links but one on one particular column. 
In both cases, roughly half of the links belongs to the maximal tree, reducing the number of samples we need to take.

Note that this holds for any gauge invariant state which can be expanded in the configuration basis; it is entirely unrelated to the efficiency of calculating the integrands.
The efficiency of computing the integrands is achieved by using the gauged Gaussian PEPS discussed here. 
The simplifying result due to gauge fixing is due only to gauge invariance.

\section{Summary and Conclusions}
\label{sec:conclusions}
In this paper, we presented a unified framework for constructing a gauged Gaussian PEPS ansatz useful for studying lattice gauge theory. 
The ansatz satisfies a built-in entanglement area law, and allows for the efficient computation of expectation values of observables using covariance matrices.

We improved on the versions of the ansatz introduced in previous work and generalized it in several ways: multiple types of virtual degrees of freedom, relating to the coupling to matter or its absence; multiple flavors of physical matter and chemical potentials; and a generalized choice of the initial gauge field configuration.
We also included some details on the numerical implementation of the algorithm that allow for more efficient calculations. 
Of particular interest is the division of the ansatz into layers and flavors, which allows the calculation of observables relative to each layer separately, with the intermediate results then easily combined to give the full expectation value, as well as the possibility of gauge fixing to significantly reduce the space of gauge configurations that must be considered. 
These allows for a significant reduction in computational load.

In previous works~\cite{emonts_variational_2020,emonts_finding_2023}, the ansatz described here was used to find the ground state of pure gauge $\mathbb{Z}_3$ and $\mathbb{Z}_2$ lattice gauge theories. 
In future work we plan to demonstrate similar results when including physical matter, including a system known to suffer from the sign problem. 
Ultimately, this ansatz can be applied to the study of non-Abelian lattice gauge theories and the difficult but fascinating phenomena they exhibit.

\section*{Acknowledgements}
U.B. acknowledges support from the Israel Academy of Sciences and Humanities through the Excellence Fellowship for International Postdoctoral Researchers. E.Z. acknowledges  the support of the Israel Science Foundation (grant No. 523/20). Co-funded by the European Union (ERC, OverSign, 101122583). Views and opinions expressed are however those of the author(s) only and do not necessarily reflect those of the European Union or the European Research Council. Neither the European Union nor the granting authority can be held responsible for them.

P.E. acknowledges the support received by the Dutch National Growth Fund (NGF), as part of the Quantum Delta NL programme and through the NWO-Quantum Technology programme (Grant No. NGF.1623.23.006). 
The views and opinions expressed here are solely those of the authors and do not necessarily reflect those of the funding institutions.

\bibliography{references.bib}

\appendix

\section{The Gaussian Formalism and Our Ansatz}
\label{sec:covariance}

In this appendix we shall briefly review some of the basics of the \emph{fermionic} Gaussian formalism~\cite{bravyi_lagrangian_2005}. 
We leave the details for bosons to future work.

Let $\{a_l^{\dagger}\}^N_{l=1}$ be a set of fermionic creation operators, from which, together with the corresponding annihilation operators $\{a_l\}^N_{l=1}$, we can construct a set of $2N$  Majorana modes,
\begin{equation} \label{eq:majorana-def}
    \gamma_l^{(1)}=a_l + a^{\dagger}_l , \quad \gamma_l^{(2)}=i\left(a_l - a^{\dagger}_l\right),
\end{equation}
which are all hermitian and square to the identity. 
If we relabel the Majorana modes with a new index $a \in \{1,...,2N\}$, we can write the Clifford algebra
\begin{equation}
    \left\{\gamma_{a},\gamma_{b}\right\}=2\delta_{ab},
\end{equation}
i.e. they anti-commute with an extra factor of $2$ compared to the modes $a_l^{\dagger}$.

For any pure fermionic Gaussian state, $\ket{\Phi}$, we can define the covariance matrix
\begin{equation} \begin{aligned} \label{eq:covmat-def}
    \Gamma_{ab}
        &= \frac{i}{2} \frac{ \expval{\comm{\gamma_a}{\gamma_b}}{\Phi}}{\braket{\Phi}} \\
        &= \langle \comm{\gamma_a}{\gamma_b} \rangle_{\Phi}.
\end{aligned} \end{equation}
The generalization for a mixed state follows. This matrix contains all the physical information contained in the Gaussian state $\ket{\Phi}$.

By definition, every covariance matrix is anti-symmetric,
\begin{equation}
\Gamma_{ba}=-\Gamma_{ab}.
\end{equation}
For pure states, 
\begin{equation}
\Gamma^2=-\id.
\end{equation}

Recall the BCS form of equation~\eqref{eq:bcs-form},
\begin{equation} 
\exp \bigg( \mathcal{M}_{mn}
a^{\dagger}_{m} a^{\dagger}_{n}
 \bigg) \ket{\Omega}. 
\end{equation} 
Any pure fermionic Gaussian state state can be written in this form. A closed formula for the covariance matrix, in terms of the matrix coupling the creation operators in the exponential (e.g. the $\mathcal{T}$ tensor in the $A$ operators in the main text, and $\mathcal{M}$ here) may be found in an appendix of~\cite{emonts_finding_2023}, and is given by
\begin{equation} \label{eq:bcs-to-covmat}
    \Gamma^D = i
        \begin{pmatrix}
            -\mathcal{M}^-\mathcal{M}  & \frac{1}{2} \mathcal{M}^-(1 + \mathcal{M} \bar{\mathcal{M}}) \\
            - \frac{1}{2} \mathcal{M}^-(1 + \bar{\mathcal{M}}\mathcal{M} ) & \mathcal{M}^-\bar{\mathcal{M}}
        \end{pmatrix}
\end{equation}
where $\bar{\mathcal{M}}$ is the complex conjugate of $\mathcal{M}$ and $\mathcal{M}^- = (1 - \mathcal{M}\bar{\mathcal{M}})^{-1}$. 
The $D$ superscript serves to emphasize that this result is stated in terms of Dirac modes $\{a_l^{\dagger}\}^N_{l=1}$. 
It can be converted to the covariance matrix of equation~\eqref{eq:covmat-def} through the relations in equation~\eqref{eq:majorana-def}, as discussed more below.

With this background, we show how to construct the covariance matrices of our ansatz $\ket{\Psi}$, and its parts.
Our state is composed out of several layers, all of the same form, given in equations~\eqref{psi2F} and~\eqref{psiIdef-per-layer}, as 
\begin{equation} 
\label{eq:layer-agnostic-state}
\bra{\Omega_\text{v}} \prod_{\site{x},k}
    w^ \dagger(\site{x}, k) 
    \mathcal{U}_G [\mathcal{G}]  \prod_{\site{x}} A(\site{x})  
    \ket{\Omega}.
\end{equation}
As discussed above, this is either a number (for layers with no physical modes) or a state (for layers with physical modes). In what follows, all quantities should be indexed to the appropriate layer, but we leave this out to avoid clutter (though we use the subscripts $I$ and $II$ to differentiate layers with/without physical matter when necessary).

We start by defining the state
\begin{equation}
\begin{aligned}
    \ket{A}&=\prod_{\site{x}} A(\site{x}) \ket{\Omega}\\
\end{aligned}
\end{equation}
and denote its density matrix by
\begin{equation}
    \rho^A =\ket{A}\bra{A}.
\end{equation}

The state $\ket{A}$ is a Gaussian product state, that is, a product of Gaussian states defined separately at each site $\site{x}$. Translation invariance implies that all of these local states will have the same covariance matrix, $M_0$, and therefore the covariance matrix of $\ket{A}$ is
\begin{equation}
    M = \bigoplus_{\site{x}} M_0,
\end{equation}
a direct sum of identical local covariance matrices. We can reorder the  modes of $\ket{A}$ such that those corresponding to physical fermions (if there are any) appear first, and write its covariance matrix in the form:
\begin{equation}
M =
\left( {\begin{array}{cc}
		M_{A} & M_{B} \\
		-M_{B}^T & M_D \\
\end{array} } \right),
\end{equation}
where $M_{A}$ is the block of correlations between physical fermions, $M_D$ between virtual ones and $M_B$ between physical and virtual. For layers with no physical modes, we simply have that $M = M_D$.

We also define the state
\begin{equation}
    \ket{B} = \prod_{\site{x}, k} w(\site{x}, k) 
    \ket{\Omega},
\end{equation}
and its density matrix as
\begin{equation}
    \rho^B =\ket{B}\bra{B}.
\end{equation}
The covariance matrix of this state, containing only virtual modes, is denoted by $\Gamma_{\text{in}}$, and can be computed by hand once, since it does not depend on the parameters of the state contained in $\mathcal{T}$.

Next we define the gauged link states,
\begin{equation}
\rho^B(\mathcal{G}) = 
\mathcal{U}^{\dagger}_G\left[\mathcal{G}\right]
    \rho^B
    \mathcal{U}_G\left[\mathcal{G}\right],
\end{equation}
and denote its covariance matrix by $\Gamma_{\text{in}}\left(\mathcal{G}\right)$. We show how to calculate this covariance matrix below.

We first consider the norm for layers without physical matter, for which the expression~\eqref{eq:layer-agnostic-state} is simply a number, $\psi_I$. Note that in the main text we use $\psi_I$ to refer to the product of all layers without matter; here we are considering just one.
Using equation~\eqref{psiIdef}, we can express it as
\begin{equation} \label{eq:psi_I_norm}
    \left|\psi_I \left(\mathcal{G}\right)\right|^2 =
    \text{Tr}\left[
    \mathcal{U}^{I \dagger}_G\left[\mathcal{G}\right]
    \rho^B_{I}
    \mathcal{U}^{I}_G\left[\mathcal{G}\right]
    \rho^A_{I}
    \right]
    \equiv
\text{Tr}\left[
    \rho^B_I\left(\mathcal{G}\right)
    \rho^A_I
    \right],
\end{equation}
which is simply the overlap of two Gaussian density matrices. This can be expressed by a closed formula in terms of their covariance matrices~\cite{bravyi_lagrangian_2005,zohar_combining_2018}, as
\begin{equation}
    \left|\psi_{I} \left(\mathcal{G}\right)\right|^2 =    
    \sqrt{\det \bigg(\frac{1-\Gamma_{\text{in}}\left(\mathcal{G}\right)M_D}{2} \bigg)},
    \label{eq:norm_gaussian_state_ferm_I}
\end{equation}
where $\Gamma_{\text{in}}(\mathcal{G})$ and $M_D$ are of course the matrices corresponding to the layer under consideration.

Now we consider layers with matter, and denote the state for one such layer as
$\ket{\psi_{II}\left(\mathcal{G}\right)}$, denote its density matrix by $\rho_{\text{out}}\left(\mathcal{G}\right)$, and its covariance matrix by $\Gamma_{\text{out}}\left(\mathcal{G}\right)$. 
This density matrix can be expressed as
\begin{equation}
\rho_{\text{out}}\left(\mathcal{G}\right) = 
\text{Tr}_{\text{v}}\left[
    \rho^B_{II}\left(\mathcal{G}\right)
    \rho^A_{II}
    \right],
\end{equation}
where $\text{Tr}_{\text{v}}$ refers to a partial trace over the virtual modes. Using Gaussian mapping techniques~\cite{bravyi_lagrangian_2005,kraus_fermionic_2010}, we can write the  covariance matrix of $\ket{\psi_{II}\left(\mathcal{G}\right)}$ as
\begin{equation}
 \Gamma_{\text{out}}\left(\mathcal{G}\right) = M_A + 
 M_B
 \left(M_D-\Gamma_{\text{in}}\left(\mathcal{G}\right)\right)^{-1}
 M^T_B.
\end{equation}
This covariance matrix is useful, for example, for computing mesonic expectation values.

Finally, to obtain the squared norm of $\ket{\psi_{II}\left(\mathcal{G}\right)}$, we can use the overlap formula again, by taking a full trace, resulting in
\begin{equation}
\begin{aligned} \label{eq:psi_II_norm}
\braket{\psi_{II}\left(\mathcal{G}\right)}
&=
\text{Tr}\left[
    \rho^B_{II}\left(\mathcal{G}\right)
    \rho^A_{II}
    \right]\\
    &=\text{Tr}\left[
    \left(\rho^B_{II}\left(\mathcal{G}\right)\otimes\id_{\text{physical}}\right)
    \rho^A_{II}
    \right]\\
    &\propto
    \sqrt{\det \bigg(\frac{1-\Gamma_{\text{in}}\left(\mathcal{G}\right)M_D}{2} \bigg)},
\end{aligned}
\end{equation}
where we have embedded $\rho^B_{II}\left(\mathcal{G}\right)$ in the full Hilbert space with the identity matrix for the physical modes. The corresponding density matrix will be a direct sum of a zero block (for the physical modes) and $\Gamma_{\text{in}}\left(\mathcal{G}\right)$. It is therefore unsurprising that the previous result holds, and equations~\eqref{eq:psi_I_norm} and~\eqref{eq:psi_II_norm} match.

It remains to be shown how to calculate $\Gamma_{\text{in}}(\mathcal{G})$ when given $\Gamma_{\text{in}}$ and $\mathcal{G}$.
To do so, we introduce the vector $C$, which contains, in order, all of the virtual annihilation operators, followed by all the virtual creation operators.
Note that when we gauge, following equation~\eqref{UG2}, we get that
\begin{equation}
    \mathcal{U}^{\dagger}_G\left[\mathcal{G}\right]
    C_i
    \mathcal{U}_G\left[\mathcal{G}\right]
    = \mathcal{V}\left(\mathcal{G}\right)_{ij}C_{j},
\end{equation}
where $\mathcal{V}\left(\mathcal{G}\right)_{ij}$ is a unitary matrix, obtained as a direct sum of the transformations on each link separately; only the outgoing links are gauged following the gauging rules from above. The modes on the incoming links, as well as to those corresponding to outgoing links on the maximal tree $\mathbf{T}$ if gauge fixing is employed (see section \ref{sec:gauge-fixing}), are untransformed by the gauging, and the corresponding blocks of $\mathcal{V}\left(\mathcal{G}\right)_{ij}$ are simply identities.

It is convenient to work with Majorana modes, so we define the matrix $S$ as that transfers $C_i$ to the corresponding Majorana modes 
$\gamma_{\alpha}$, as
\begin{equation}
\gamma_\alpha = S_{\alpha,i}C_i.
\end{equation}
where
\begin{equation}
    S = 
    \left( {\begin{array}{cc}
		\id & \id \\
		i\id & -i\id \\
\end{array} } \right).
\end{equation}
Note that $S^{-1}=\frac{1}{2}S^{\dagger}$.
    
Therefore, the gauging rule for Majorana modes is given by
\begin{equation}
\begin{aligned}
    \mathcal{U}^{\dagger}_G\left[\mathcal{G}\right]
        \gamma_\alpha
        \mathcal{U}_G\left[\mathcal{G}\right] 
    &= 
    S_{\alpha,i} \mathcal{V}\left(\mathcal{G}\right)_{ij} C_{j}
    \\ &=
    \frac{1}{2}S_{\alpha,l}\mathcal{V}\left(\mathcal{G}\right)_{ij}S^{\dagger}_{j\beta}\gamma_\beta
    \\&\equiv O\left(\mathcal{G}\right)_{\alpha\beta}\gamma_\beta,
\end{aligned}
\end{equation}
where $O\left(\mathcal{G}\right)$ is a block diagonal orthogonal matrix, with identity blocks corresponding to the modes of ungauged legs. Therefore,
\begin{equation}
 \Gamma_{\text{in}}\left(\mathcal{G}\right)_{\alpha\beta}
       = O\left(\mathcal{G}\right) \Gamma_{\text{in}} O^T\left(\mathcal{G}\right).
\end{equation}

Using these norms and covariance matrices, one can calculate the expectation value of any observable of interest, as discussed in section~\ref{sec:algorithm}.
As described there, the norms and covariance matrices must be calculated at each Monte Carlo step, for each $\mathcal{G}$.
However, $\rho^A$ does not depend on $\mathcal{G}$.
Therefore in each Monte-Carlo step, only $\Gamma_{\text{in}}\left(\mathcal{G}\right)$, corresponding to $\rho^B$, must be evaluated, and this can be made more efficient using local updates, as detailed in appendix~\ref{sec:local_updates}.

\section{Monte Carlo with Local Updates}
\label{sec:local_updates}
The computation of observables with the GGPEPS ansatz relies on the sampling of gauge configurations.
In Monte Carlo, there are generally two options to sample new configurations.
The update can either be local or global~\cite{wolff_collective_1989}.
In a local update, a fixed number of gauge fields are updated.
With increasing system-size, the fraction of gauge fields that are updated decreases.
This is different for global updates~\cite{wolff_collective_1989,swendsen_nonuniversal_1987}, for which the number of changed gauge fields scales with the system size.

In general, global updates are more computationally demanding per update step but lead to a better ergodicity and a faster convergence.
However, finding a global update scheme for a given system demands fine-tuned updates.
In this work, we use local updates changing the gauge field configuration underlying the covariance matrix $\Gamma_{\text{in}}(\mathcal{G})$ locally.

The local updates enable a speed-up in the computation of the weight (unnormalized probability) for the next configuration and the observables.
These optimizations use the Woodbury matrix identity~\cite{higham_accuracy_2002} and the determinant lemma~\cite{harville_matrix_1997}.
During the Monte Carlo sampling procedure new gauge field configurations are accepted or rejected depending on their weight, computed as the norm of the state $\ket{\psi(\mathcal{G})}$.
This weight takes the form of a matrix determinant, as seen in equations~\eqref{eq:norm_gaussian_state_ferm_I} and~\eqref{eq:psi_II_norm}.

The matrix determinant lemma helps to compute the determinant of a matrix after a local change.
The determinant of a matrix $M$ with a local change $C$ is computed as
\begin{equation}
  \det(M+UCV\tran) = \det(C\inv+V\tran M\inv U) \det(C) \det(M),
  \label{eq:det_lemma}
\end{equation}
where $U$ and $V$ are meant to position the update $C$ with respect to the larger matrix $M$.
Both $U$ and $V$ are either $0$ or blocks of the identity matrix.
In the case of GGPEPS, $M=(D\inv-\gammain(\mathcal{G}))$, and $C$ is the local change of the block-diagonal matrix $\gammain(\mathcal{G})$.
The matrices $U$ and $V$ position the change at the block corresponding to the changed gauge field.
However, to compute the determinant, we need to know the inverse of $M$, i.e. $(D\inv-\gammain(\mathcal{G}))^{-1}$ in the case of GGPEPS.

Conveniently, the Woodbury formula can update the inverse of a matrix $M$ after a local change $C$ given the determinant before the change
\begin{equation}
  (M+UCV)\inv=M\inv+M\inv U(C\inv+VM\inv U)\inv V M\inv.
  \label{eq:woodbury}
\end{equation}
For a local Monte Carlo update, the size of the updated submatrix is $C$ is constant.
Thus, we can apply the Woodbury formula, which provides an efficient update algorithm for the inverse of a matrix which is only slightly altered.
The drawback is only that the inverse matrix has to be stored in order to perform the next update.

At each step of the Monte Carlo algorithm, the inverse is updated according to the determinant of the previous step.
Then, the determinant is updated using the inverse including the new gauge configuration.
In total, we can update the determinant for the norm, the inverse $(D\inv-\gammain(\mathcal{G}))^{-1}$ and the inverse for $(D-\gammain(\mathcal{G}))^{-1}$ used for the computation of $\Gamma_{\text{out}}(\mathcal{G})$.
This tracking of determinants and inverses enables updates in $\mathcal{O}(N^2)$ instead of $\mathcal{O}(N^3)$.
Determinants and inverses are only computed from scratch once, for the first gauge field configuration.

A central disadvantage of local update is the high auto-correlation between successive update steps, leading to larger errors in the Monte Carlo estimators.
By changing only a single gauge field, a large portion of the system remains unchanged.
Given the tracking of determinants and matrix inversions, updates are usually much faster to compute than certain observables.
Thus, by updating multiple times between measurement steps, the statistical error on observables can be reduced since the auto-correlation between successive measurements decreases.
\end{document}